\documentstyle[aasms,psfig,epsf]{article}

\begin{document}

\def\beq{\begin{equation}}
\def\eeq{\end{equation}}
\def\eq{equation}
\def\eqs{equations}
\def\t{\theta}
\def\sb{surface brightness}
\def\Nlz{N(>l,z)}
\def\Nltz{N(>l, >\t, z)}
\def\nlz{n(l,z)}
\def\nltz{n(l, >\t, z)}
\def\Lm{4\pi d_L^2 l}
\def\biso{b_{\rm iso}}
\def\Biso{B_{\rm iso}}
\def\Z4{(1+z)^4}
\def\riso{r_{\rm iso}}
\def\xiso{x_{\rm iso}}
\def\yiso{B_0/B_{\rm iso}}
\def\tiso{\theta_{\rm iso}}
\def\Liso{L_{\rm iso}}
\def\liso{l_{\rm iso}}
\def\lt{\pi \t^2 \biso}
\def\bz{\Z4 b}
\def\Lmint{L_{{\rm min}, \t}}
\def\Lminl{L_{{\rm min}, l}}


\title {NEW AND OLD TESTS OF COSMOLOGICAL MODELS AND EVOLUTION OF GALAXIES}

\author{Vah\'{e} Petrosian}

\affil{Space Telescope Science Institute, Baltimore, MD 21218} 
\affil{Center for Space Science and Astrophysics, Stanford University, 
Stanford, CA 94305}



\begin{abstract}


We describe the classical cosmological tests, 
such as the Log$N$-Log$S$, redshift-magnitude and angular diameter tests,
and propose some new tests of the evolution of galaxies and the
universe. Most analyses of these tests treat the problem in terms of a 
luminosity function and its evolution. The main thrust of this paper is to
show that this is inadequate and can lead to a incorrect conclusions when 
dealing with high redshift sources. We develop a proper treatment in three 
parts. 

In the first part we describe these tests based on the isophotal
values of the quantities such as flux, size or surface brightness.
We show the shortcomings of the simple point source 
approximation based solely on the 
luminosity function and consideration of the flux limit. We emphasize
the multivariate nature of the problem and quantify the effects of 
other selection biases due to the \sb\ and angular size limitations. 
In these considerations the surface
brightness profile, and the distribution of the basic parameters
describing it, play a critical role in modeling of the problem.
In general, in the isophotal scheme the data 
analysis and its comparison with the model
predictions is complicated. In the second part we 
show that considerable simplification
is achieved if these test are carried out in some sort of metric scheme, 
for example that suggested by Petrosian (1976). This scheme, however,
is limited to well resolved sources.
Finally, we describe the new tests and compare them to the traditional tests
demonstrating the observational and modeling ease that they provide. 
These new procedures, which can use the data to a fuller extent
than the isophotal or metric based tests,
amount to simply counting the pixels
or adding their intensities as a function of the surface
brightness of all galaxies instead of dealing with surface brightnesses, 
sizes and fluxes (or magnitudes) of individual galaxies.
We also show that the comparison of the data with the theoretical 
models of the distributions
and evolution of galaxies has the 
simplicity of the metric test and utilizes the data as fully as the 
isophotal test.

{\it Subject headings}: cosmology:theory--galaxies:evolution--galaxies:
luminosity function--galaxies:photometry.

\end{abstract}

\section{INTRODUCTION}

Galaxies and other extragalactic sources provide the most direct
means of studying  evolution in the universe. This is done using
the classic cosmological tests
such as the angular diameter-redshift-magnitude relations
or the source counts (also known as the Log$N$-Log$S$) test.
(For a general description of these tests see, e.g. Weinberg 1972.)
These tests, which rely primarily on the distribution of the
magnitudes or fluxes of the sources, have 
had limited success in determining the cosmological parameters and/or
the evolution of galaxies. There are two fundamental reasons for
this failure. The first is the well-known difficulty of disentangling the
evolution of the sources (see e.g. Tinsley 1968 or Tinsley \& Gunn 1976) 
from the evolution of the universe (Weinberg 1972).
As a result, over the years,
the focus of such studies has been shifted from
the determination of the cosmological parameters to the evolution
of galaxies in different assumed cosmological models.
The second difficulty arises from the fact that galaxies are extended (i.e.
resolved) sources and there is ambiguity in defining proper
magnitudes (or luminosities $L$ and fluxes $l$)
and diameters. In addition  the samples of sources are not merely
limited by their fluxes (or magnitudes) but  
there exist other selection biases or data truncations due
to surface brightness or size limitations. These aspects of the problem are 
usually ignored. This may be an acceptable approximation for high surface 
brightness sources at low redshifts, but it is woefully inadequate when
dealing with data at high redshifts extending to low surface brightness 
sources. The main purpose of this paper is to present a proper analysis
of the various observational biases that are encountered in this process.
There are two ways one can carry out this task. From the original data sample 
one can select a subsample with fewer and simpler biases (as we do in 
\S {\bf 3}),
or one may correct the model expectation fully for all known selection biases,
which is the approach we take in the rest of the paper. The first method 
is more appropriate for tests dealing with the moments of the distributions
such as flux-redshift or size-redshift relations. The second method is
preferable when dealing with the various source counts and uses 
all of the valuable data.

The bias 
due to the magnitude or flux limit is accounted for by various means.
The most common practice is to use {\em isophotal} values, i.e. the values
of these quantities up to or at some limiting apparent surface
brightness $\biso$. However, because of the rapid
decline of the apparent surface brightness $b$ (defined as flux per
unit angular area of a resolved source) with redshift $z$ (see, e.g. Tolman
1934), 
\beq\label{basic}
	b=B/\Z4, 
\eeq
where $B$ is the intensity or the absolute surface brightness
at the source, the biases due to the surface 
brightness and size limits of the observations become important
at high redshifts and/or for low surface brightness sources.
These effects are often ignored or are dealt with indirectly
by using a limited portion of the available data.

The corrections required for these effects, sometimes referred to as
aperture corrections, inevitably require the
knowledge of the surface brightness profile
\beq\label{profile}
	B(r)=B_0f(r/r_0)\ \ {\rm with}\ \  f(0)=1,
\eeq
and the distributions of the
central surface brightness $B_0$, the characteristic or core radius
$r_0$, and other parameters $\alpha_i$ defining the profile $f(x)$.
For example, $\alpha=1\, {\rm or}\, 1/4$ for disks or spheroids, respectively,
where the profile is described by the simple 
relation ${\rm ln}f(x)=-x^\alpha$. 
Early examples of methods to correct the redshift magnitude relation 
for the aperture effect were described by
Sandage (1972), using an iterative procedure,
and by Gunn \& Oak (1975), assuming a fiducial cosmological model.
These authors used empirical relations for 
 the luminosity within the radius $r$, 
\beq\label{lumin}
	L(r)=(4\pi )\pi r_0^2B_0F(r/r_0)\ \ {\rm with} \ \  
	F(t)=\int_0^t 2xf(x)dx.
\eeq
It was shown by Petrosian (1976) (P76 hereafter) that
these corrections can be carried out more directly.
It was also shown in P76 that one can separate the evolution of the 
surface brightness $B_0$ from the evolution of universe, and
can avoid some of the above difficulties by dealing with the 
angular sizes and magnitudes up to and within a 
``proper metric'' radius $r_p$ obtained from a specified
value of the quantity 
\beq\label{eta}
	\eta={{F(x)} \over {x^2f(x)}}=2 {{d{\rm ln}r} \over 
	{d{\rm ln} L(r)}} \ \ {\rm with} \ \ x={r\over r_0},
\eeq
which is equal to the ratio of the average \sb\ within $r$
to the \sb\ at $r$.

The above equations describe the source brightness profile and its basic
properties in terms of two convenient parameters; the central \sb\ $B_0$ 
and core radius $r_0$. These parameters are not easily accessible 
to observations and their relative values for different values of $\alpha$
are difficult to interpret. This difficulty can be
overcome if we transfer the above relations to 
observationally more meaningful parameters. One commonly used such sets 
of parameters
is the effective radius and \sb. 
The effective values refer to the radius containing half the 
total luminosity $L$ which means $F(r_{\rm eff}/r_0)=0.5F(\infty)$.
%
%
The ratios of the effective to central values are: $B_{\rm eff}/B_0 = 0.189, 
2.54\times10^{-2}, 3.45\times10^{-3}$  and
$4.66\times10^{-4}$,  and $r_{\rm eff}/r_0 = 1.66, 13.5, 1.82\times10^2$ 
and $3.46\times10^3$ for $\alpha = 1, 1/2, 1/3$ and 1/4, respectively. 

\begin{figure}[htbp]
\leavevmode\centering
\psfig{file=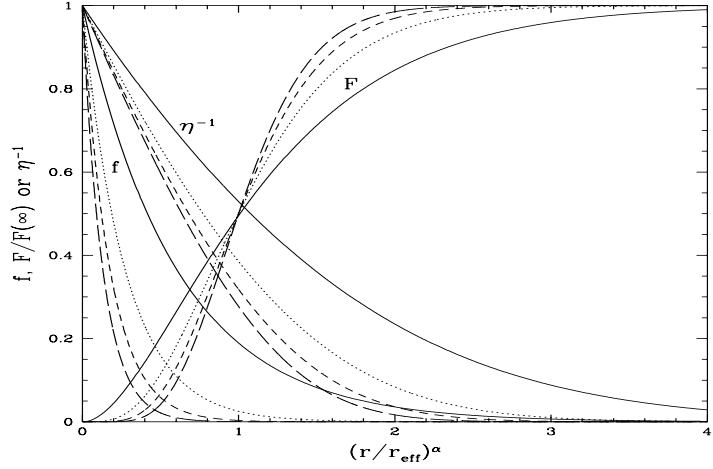,width=0.6\textwidth,height=0.4\textwidth}
\caption{Variation of \sb\ profile $f$, normalized luminosity growth curve 
$F/F(\infty)$,
and the inverse of the $\eta$ function defined in equation (\ref{eta})
versus $(r/r_{\rm eff})^\alpha$ for $\alpha=1, 1/2, 1/3$, and $1/4$ 
(solid, dotted, short dashed, and long dashed lines, respectively). 
Note that ${\rm ln}f=-(r/r_{\rm eff})^\alpha(r_{\rm eff}/r_0)^\alpha$ 
and the effective radius $r_{\rm eff}$ is defined as 
$F(r/r_{\rm eff})=0.5F(\infty)$.}
\label{profiles}
\end{figure}

Figure \ref{profiles} shows the profile $f$, the curves of growth of 
luminosity $F/F(\infty)$
and the function $\eta$  as a function of 
$(r/r_{\rm eff})^\alpha$, for the above values of $\alpha$. This shows the 
general and relative characteristics of these functions.

\begin{figure}[htbp]
\leavevmode
\centerline{
\psfig{file=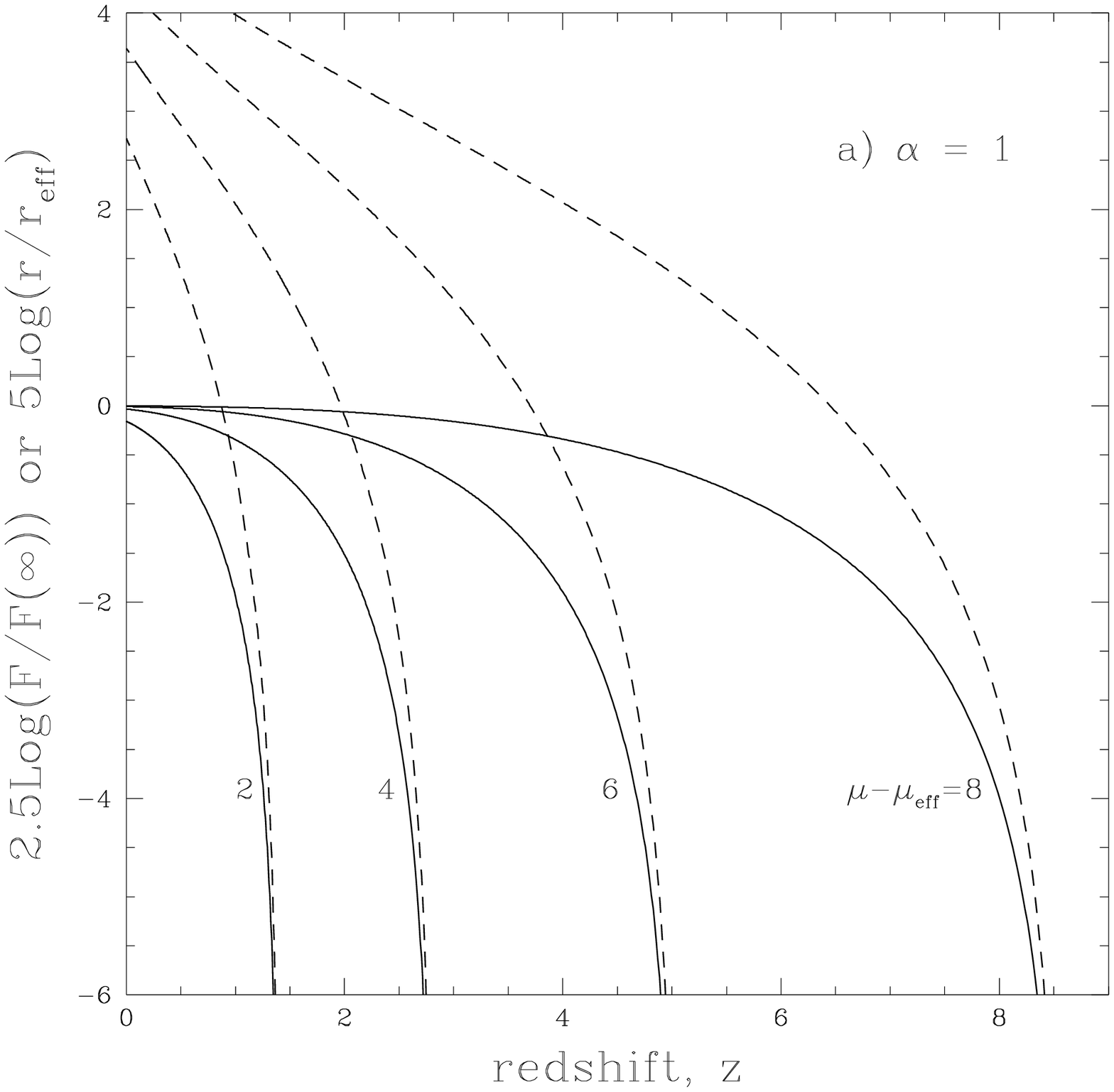,width=0.5\textwidth,height=0.5\textwidth}
\psfig{file=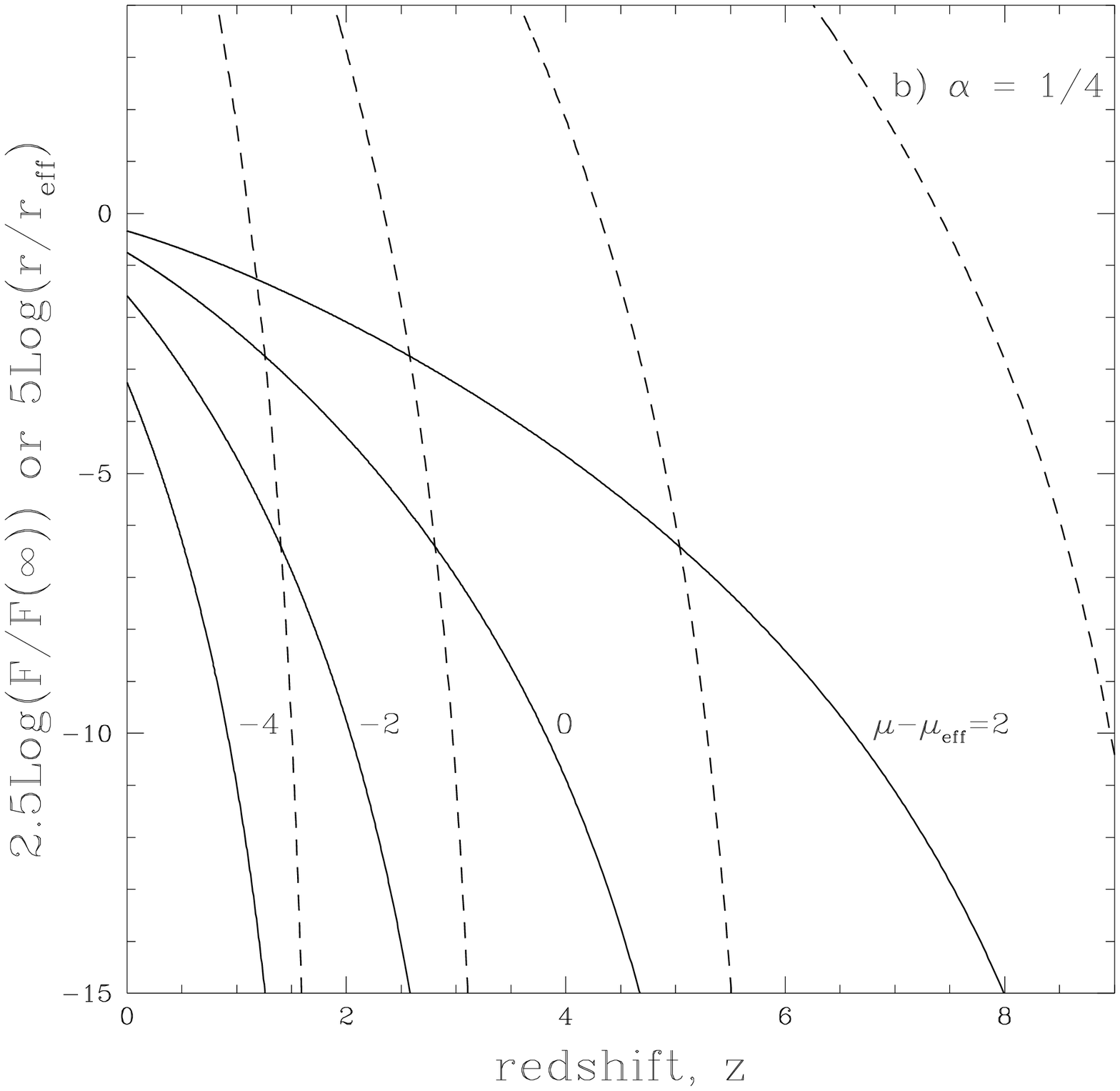,width=0.5\textwidth,height=0.5\textwidth}
}
\caption{Variation of the fraction of the luminosity (solid lines) 
and projected surface area (dashed lines),
in some arbitrary magnitude units, within a \sb\ isophot $\mu$ versus 
redshift for several values of the effective \sb\ $\mu_{\rm eff}$.
a)for an exponential (disk) profile, ${\rm ln}f=-x$;
b)for a de Vaucouleurs profile, ${\rm ln}f=-x^{1/4}$.
Note that for each value of $\mu_{\rm eff}$ the luminosity and area shrink
rapidly as we approach the maximum redshift $z_{\rm max}=(B_0/b)^{1/4}-1$,
where $B_0$ and $b$ are related to  $\mu_{\rm eff}$ and $\mu$ as described 
in the text. Also note that this decline is more pronounced for disks 
than spheroids indicating that the relative populations of sources with 
different profiles will vary strongly with redshift.
}
\label{FandAvsz}
\end{figure}

To demonstrate the effects 
of the redshift, in Figure \ref{FandAvsz}
we show the variation with redshift of the fraction of the luminosity and the 
area (expressed in magnitude units) within a \sb\ limit $b$ 
for $\alpha = 1$ and 1/4 profiles.
Instead of the surface brightnesses $B_0$ and $b$
we use the more familiar magnitudes;
$\mu = -2.5{\rm log}b\  + \ $ const., and
$\mu_{\rm eff} = -2.5{\rm log}(B_{\rm eff})\ + \ $ const.. 
As evident the observable area and luminosity vary rapidly with 
redshift, specially for low surface brightnesses and have a different 
behavior for the two profiles. This will produce a variation with redshift
of the relative abundances of disks and spheroids.
Another possible representation of the above graphs will be in terms of the
proper metric radius $r_p$ defined above. This is preferable because this 
definition of radius relies on the data within some measured isophot and not 
on the unobserved outer parts which are needed to determine the effective 
radius $r_{\rm eff}$. This  procedure will be developed further
in \S {\bf 3}.

In this paper we review several old procedures and propose 
some new ones for the study of the
evolution of galaxies, and possibly that of the universe, whereby 
instead of dealing with individual galaxies
we deal with the combined brightness of all galaxies. The new methods
simplify the data analysis enormously and are perfectly suited for modern
digitized data. In \S 2  we first give a brief description of
the proper analysis of the classical tests for isophotal quantities
that includes all the observational selection effects as well as the 
effects of the surface brightness profile, and treats the problem in 
terms of the multivariate distribution $\psi(B_0, r_0, \alpha_i, z)$
instead of the commonly used luminosity function $\phi(L, z)$.

There are several reasons for the popularity of the latter procedure.
The first is that we
have accumulated a considerable knowledge about the local luminosity function
of galaxies (see e.g. Efstathiou et al. 1988 and Lin et al. 1966) but 
little information on the distributions of the $B_0$ and $r_0$
(see, e.g. Sandage \& Perelmuter, 1991).
Secondly, until recently it was believed that that the distribution
of the surface brightness was fairly narrow (Fish, 1964; Freeman, 1970). 
However, in recent
years, because of increasing interest in low surface brightness
dwarf galaxies, some data has been accumulated on the intrinsic
distributions of these parameters indicating
broad distributions (see e.g. McGaugh 1996;
de Jong 1996; Tully \& Verheijen 1997; Impey \& Bothun 1997
and references cited there). Of course, one can
use the luminosity function by replacing the total luminosity 
$L=(4\pi)\pi r_0^2B_0F(\infty)$ [see eq. (\ref{lumin})] 
for either $B_0$ or $r_0$.
In any case a multivariate description
is required. In \S 3 we repeat the analysis of \S 2 for metric
quantities. In \S 4 we describe the new tests and their relations
to the multivariate distribution $\psi$ and the profile $f(r)$.
Finally, in \S 5 we give a brief summary. 

\section{CLASSICAL TESTS: ISOPHOTAL VALUES}

The classical test use the observed relations between the magnitude
(or flux $l$), angular size ( radius $\theta$ or area $\pi \theta^2$)
and redshift to determine the cosmological parameters and the evolution
of sources as described by the general ``luminosity'' function $\psi$. 
The cosmology is introduced via the relations 
\beq\label{ltheta}
	l(\theta)=(4\pi )\pi r_0^2B_0F(r)/\left(4\pi d_L^2(z, \Omega_i)\right) \ \ 
	{\rm and} \ \ \theta=r/d_A(z, \Omega_i),
\eeq
where $d_L$ and $d_A=d_L/(1+z)^2$ are the luminosity and the angular diameter 
distances, and $\Omega_i$ represent the cosmological parameters such as
the density parameter $\Omega$, the deceleration parameter $q_0$ or the 
cosmological constant $\Lambda$ (see, e.g. Weinberg 1972). 
All the classical tests can be described 
in terms of the observed distribution of flux, size and redshift;
$n(l,\t,z)$. For the purpose of the demonstration of the effects that
we would like to emphasize here, let us consider the cumulative source counts 
as a function of redshift, which we denote by $\Nltz$. The cumulative 
and differential counts of the so called Log$N$-Log$S$ relation is obtained by 
the integration of the above expressions over the redshift.

For simplicity, in the above relation and in what follows, we
ignore cosmological attenuation, if any,
assume either bolometric (or monochromatic) fluxes, so that we can ignore the 
K-correction, and assume spherical symmetry with brightness profile $f(r)$
independent of the wavelength. The complications due to K-correction, 
asphericity,
etc. can be easily included in the relations that follow. We will address
some aspects of these in \S 5.

\subsection{Point Sources}

The usual practice (see, e.g. Metcalfe et al. 1995 or Tyson, 1988) 
is to compare the observed cumulative and
differential distributions $\Nlz$ and
$\nlz=-\partial \Nlz/\partial l$ (and their integrals over redshift)
with that expected from models via 
the relations
\beq\label{Npoint}
	\Nlz={dV\over dz} \int_{\Lm}^\infty \phi(L,z)dL \ \ 
	{\rm and} \ \ \nlz={dV\over dz} 4 \pi d_L^2\phi(\Lm,z),
\eeq
where $V(z,\Omega_i)$ is the co-moving volume up to redshift $z$ and
$\rho(z)=\int_0^\infty \phi dL$ is the co-moving density of all sources at $z$.
Here and in what follows we assume a complete $4\pi$ sterradian sky coverage.
These expressions is what one would expect for unresolved or point sources,
where only the flux limit counts in the selection process (see, however,
a modification below in \S {\bf 2.3}). 

\subsection{Extended Sources}

For extended sources such as galaxies the 
selection process is more complex and additional corrections are required.
We now describe these selection biases.

\subsubsection{Surface Brightness Limit}

A source to be detected must have an apparent central \sb\ exceeding
the detection threshold which must be several times the
standard deviation $\sigma$ of the fluctuations in the background brightness. 
We denote this limit by $\biso$. If we ignore the image degradation due to
the finite size of the instrumental and atmospheric 
point spread function (PSF), which can be done 
if the core size 
$r_0 \gg \theta_s d_A$, and if the pixel size is less than or 
comparable to the width $\theta_s$
of the PSF,
then the surface brightness selection criterion requires that 
\beq\label{Biso}
	B_0 \geq \Biso \equiv \Z4\biso.
\eeq
However, for small sources or high redshifts the effect of the
finite size of the PSF cannot be ignored
and the selection bias is more severe than indicated by this relation. 
For a PSF$=g(\theta/\theta_s)$ 
the surface brightness is modified to
\beq\label{psfpr}
	\hat B (r)=\hat B_0\hat f (r/r_0, \t_s d_A/r_0),\ \ {\rm where} 
         \ \ \hat f =f*g 
\eeq
is the convolution of the the actual profile with the PSF. As a result,
the central surface brightness is reduced by $\xi = \hat B_0/B_0$, where
\beq\label{psfB}
	\xi (\t_s d_A/r_0)=\int_0^\infty f(\theta d_A/r_0)g(\theta/\theta_s)
	\theta d\theta\bigg/ \int_0^\infty g(\t /\t_s) \theta d\theta.
\eeq
For the purpose of illustration let us consider a box PSF with the
radial width of $\theta_s$. This reduction factor then simplifies 
and equation (\ref{Biso})
is modified to read
\beq\label{b1}
	B_0 \geq \Biso x_s^2/F(x_s) \ \ {\rm with}\ \ x_s=\theta_s d_A/r_0.
\eeq
Similar expressions can be derived for other forms of the PSF.
For $x_s \ll 1$ this reduces to equation (\ref{Biso})
but its effects become 
important for $x_s$ near unity i.e. for partially resolved and 
unresolved sources. Note that $\t_s$ is replaced by the pixel 
size $\t_{\rm pix}$ if $\t_s < \t_{\rm pix}$

\subsubsection{Size Limit}

Another criterion for selection of extended sources such as galaxies
is that their sizes must exceed some limit. One way to quantify this
is to have the isophotal angular radius
(namely, the radius where the \sb\  has dropped to the specified 
isophotal value $\biso$)
be larger than some specified size $\theta$. 
If $\theta \gg \theta_s$, then the isophotal angular radius is given by
\beq\label{tiso}
	\tiso=(r_0/d_A)f^{-1}(\Biso/B_0),
\eeq
where the function $f^{-1}$ is the inverse of the profile function $f$.
Then the selection condition $\tiso \geq \t$ is satisfied if
\beq\label{b2}
	B_0 \geq \Biso/f(\theta d_A/r_0).
\eeq
However, as $\t$ decreases toward $\t_s$ one should use the modified
profile $\hat f$ of equation (\ref{psfpr}) in place of $f$. In any case,
it is clear that
for $\theta \geq \theta_s$ this inequality will provide 
a more restrictive limit than the \sb\ limit because it requires that
more than one pixel to exceed the \sb\ limit $\biso$.
This can be demonstrated mathematically also 
by setting $\theta =\theta_s$ in the last equation and comparing it with the 
limit in equation (\ref{b1}).
The ratio of the two limits is equal to $\eta(x_s)$ which according to
equation (\ref{eta})
is greater than one, except for the unlikely event of the \sb\ increasing 
with $r$. In the opposite case when
$\t <\t_s$ one is dealing with
unresolved or point like sources in which case a size limit does not make sense.

For non-spherical sources we can follow a similar procedure by dealing
with the isophotal angular area $\omega$ (which for spherical sources is
equal to $\pi \t^2$) as the area of the sources with apparent \sb\ 
$b \geq \biso$. However, the relation of this to the surface 
brightness profile will be more complicated. For example, for 
elliptical sources with a constant ellipticity we can express
the profile $f$ as a function of the area $a/a_0$ with $a_0=\pi r_1 r_2$,
where $r_1$ and $r_2$ are the core radii along the major and minor axes.
For randomly oriented elliptical sources this will amount to replacement
of the quantity $\t d_A/r_0$ in \eq\ (\ref{b2}) by $(\omega d_A^2/a_0)^{1/2}.$
The distribution function $\psi$ now will be a function of $a_0$
and the ellipticity or the ratio $r_1/r_2$.

\subsubsection{Flux Limit}

Finally the sample of sources is subject to a flux limit. In this 
section we consider 
the flux $\liso$ within the  isophotal angular radius $\tiso$ or up to
the \sb\ limit $\biso$.
The flux limit then implies that
$\liso=L(\tiso d_A)/4\pi d_L^2$ is greater than some specified flux $l$.
Using equations  (\ref{lumin})
 and (\ref{ltheta})
 we can write this limit as
\beq\label{b3}
	B_0F(\xiso)\geq d^2_L l/(\pi r_0^2) \ \ {\rm with}
	 \ \  \xiso \equiv \tiso d_A/r_0=f^{-1}(\Biso/B_0).
\eeq
Note that the left hand side of this inequality is independent of $r_0$.
However, if $\tiso\not\gg\t_s$, then one should replace the profiles
$f$ and its integral the luminosity growth curve $F$ 
by the corresponding values, $\hat f$ and $\hat F$, 
modified by the PSF. 
In this case the above relation becomes more complex with the 
involvement of the additional variable $\t_sd_A/r_0$ (see \S {\bf 2.4} below).

\subsubsection{Combined Limits}

\begin{figure}[htbp]
\leavevmode
\centerline{
\psfig{file=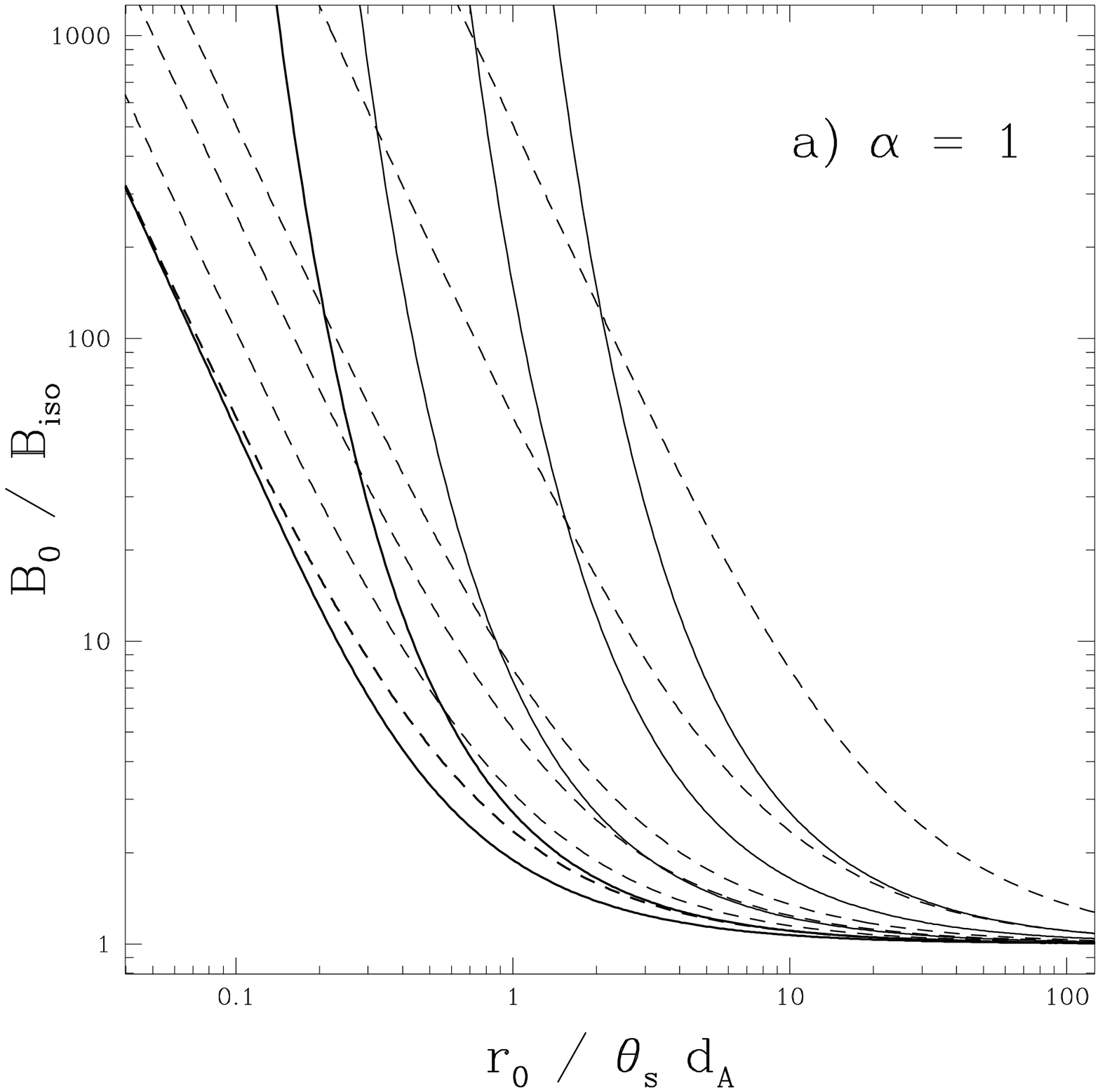,width=0.5\textwidth,height=0.5\textwidth}
\psfig{file=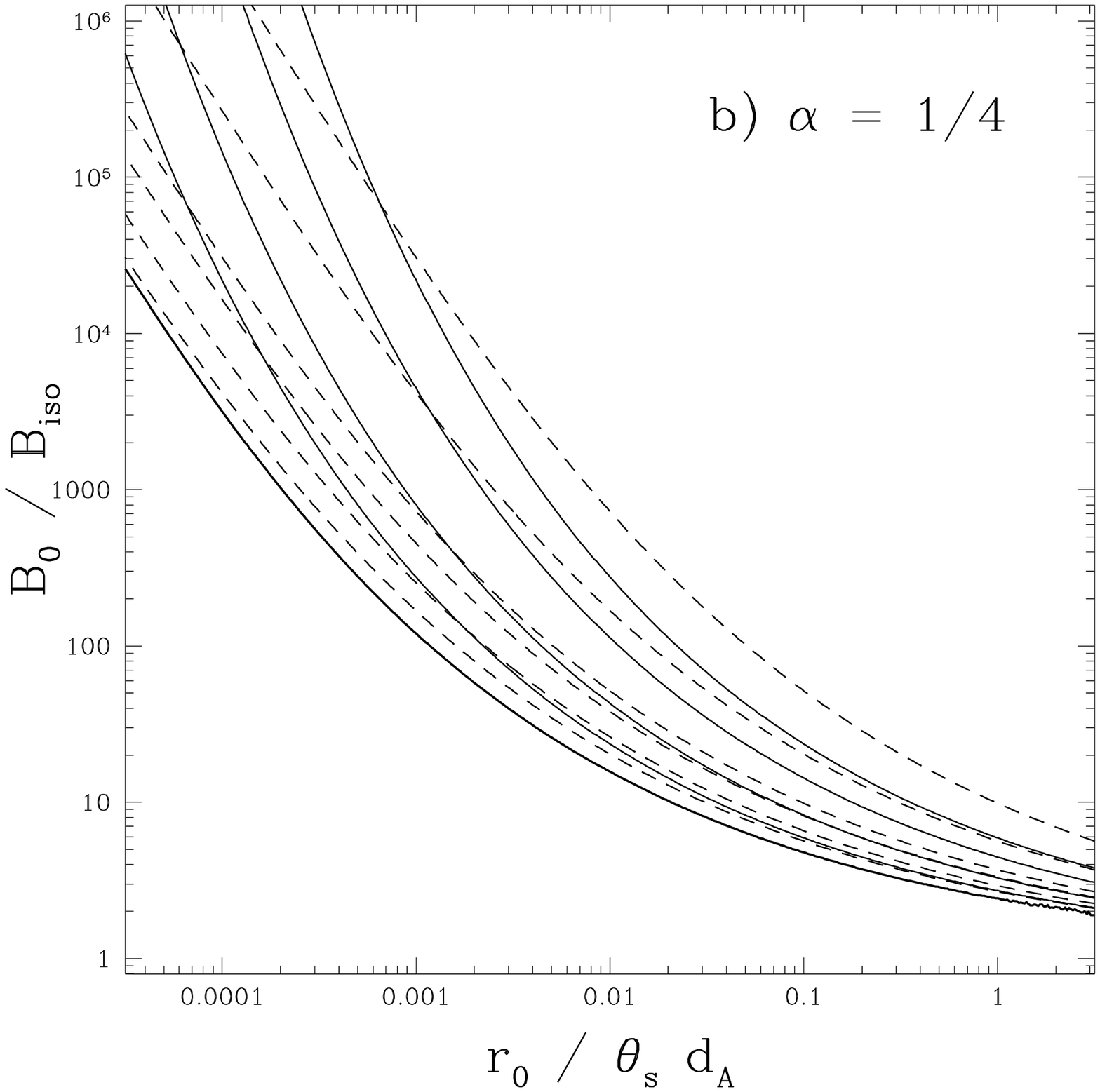,width=0.5\textwidth,height=0.5\textwidth}
}
\caption{
Central \sb\ $B_0$ in units of $\Biso=\biso \Z4$ versus the size $r_0$ in units
of $\t_s d_A$ depicting the \sb\ limit (the heavy or lowest
solid line); the angular size limit  
for $\t/\t_s=1,2,5 \ {\rm and}\ 10$ (thin solid lines);
and the flux limit for  $l/\pi \t_s^2 \biso= 1,2,5,10,100 \ {\rm and}\ 1000$ 
(dashed lines).
a)for an exponential (disk) profile, ${\rm ln}f=-x$;
b)for a de Vaucouleurs profile, ${\rm ln}f=-x^{1/4}$.
Note that the solid and dashed lines crosss each other where the 
limiting factor changes from size to flux. The region to the right of 
the appropriate sold line before these intersections and to the right 
of the appropriate dashed line after the intersections are accessible 
to observation. Note also that, because $\Biso$ and $d_A$ increase with
redshifts, the observable region shrinks systematically, moving towards
the upper right hand corner, as redshift increases.
}
\label{B0r0}
\end{figure}

Thus, for a given value of $\t, l, \biso$ and $\t_s$ the above three 
inequalities determine the region of the $B_0-r_0$ plane that is accessible
at these particular conditions. This region varies with redshift 
becoming smaller at higher  redshifts. The redshift dependences
are hidden in $\Biso$ and $d_A$. Figure \ref{B0r0} show the three
boundary conditions obtained by the equality sign in
equations (\ref{b1}), (\ref{b2}) and (\ref{b3})
for the exponential, $\alpha=1$, and de Vaucouleurs, $\alpha=1/4$, 
profiles.
We plot $B_0/\Biso$ versus $r_0/(\t_s d_A)$, which is valid at 
all redshifts. The lowest (heavy) solid line shows the truncation 
due to the \sb\ limit. The lighter 
solid lines show the effects of the size limit for several values of
$\t/\t_s$. The dashed lines show the truncation due to the 
flux limit for different values 
of the ratio  $l/(\pi \t_s^2 \biso)$. 
Sources lying in the region above all three
lines are the ones which satisfy all the selection criteria. 
It is clear that, as long as $\t>\t_s$, or $l>\pi\t_s^2\biso$,
which obviously will be the case for resolved sources, 
the surface brightness limit due to the PSF,
described by equation (\ref{b1}),
is never important. However, both size and
flux limits could be important depending on the relative values of the 
observational limits. For larger values of the ratio $l/(\pi \t^2 \biso)$
the flux limit provides the major constraint. In the opposite case
the size limit becomes more important, and as evident from the above
figures for $l=\pi \t^2 \biso$ only the size limit is relevant.

\subsubsection{Extended Source Counts}

We can now relate the observable $\Nltz$ to the distribution function 
$\psi(B_0,r_0,z)$ 
by the integration of the latter over the accessible region as determined
by the observational limits. For the general case this gives
\beq\label{Next}
	\Nltz={dV\over dz}\left(\int^{B_{0,cr}}_{\Biso} dB_0
        \int^\infty_{r_{0,1}} dr_0 \psi(B_0,r_0,z) +
        \int^\infty_{B_{0,cr}} dB_0
        \int^\infty_{r_{0,2}} dr_0 \psi(B_0,r_0,z) \right).
\eeq
Here $r_{0,1}$ and $r_{0,2}$ are obtained by solving equations (\ref{b3}) 
 and (\ref{b2})
for $r_0$ in terms of $B_0$ and other observables, and 
\beq\label{Bcr}
	B_{0,cr}= \Biso/f \left(\eta^{-1}\left(l/(\pi \t^2 \biso) \right)
	\right),
\eeq
is the intersection point of the the two boundary conditions
described by equations (\ref{b2})
and (\ref{b3}) or the intersection of a 
solid and a dashed line in Figure \ref{B0r0}; 
$\eta^{-1}$ is the 
inverse function of the function $\eta$ defined in equation (\ref{eta}).
Note that as stated above for $l=\pi \t^2 \biso$ this critical value of 
\sb\ becomes equal to $\Biso$, the first double integral in 
the left hand side of \eq\ (\ref{Next}) vanishes and we
are left with the size limited part of this expression only.

\begin{figure}[htbp]
\leavevmode
\centerline{
\psfig{file=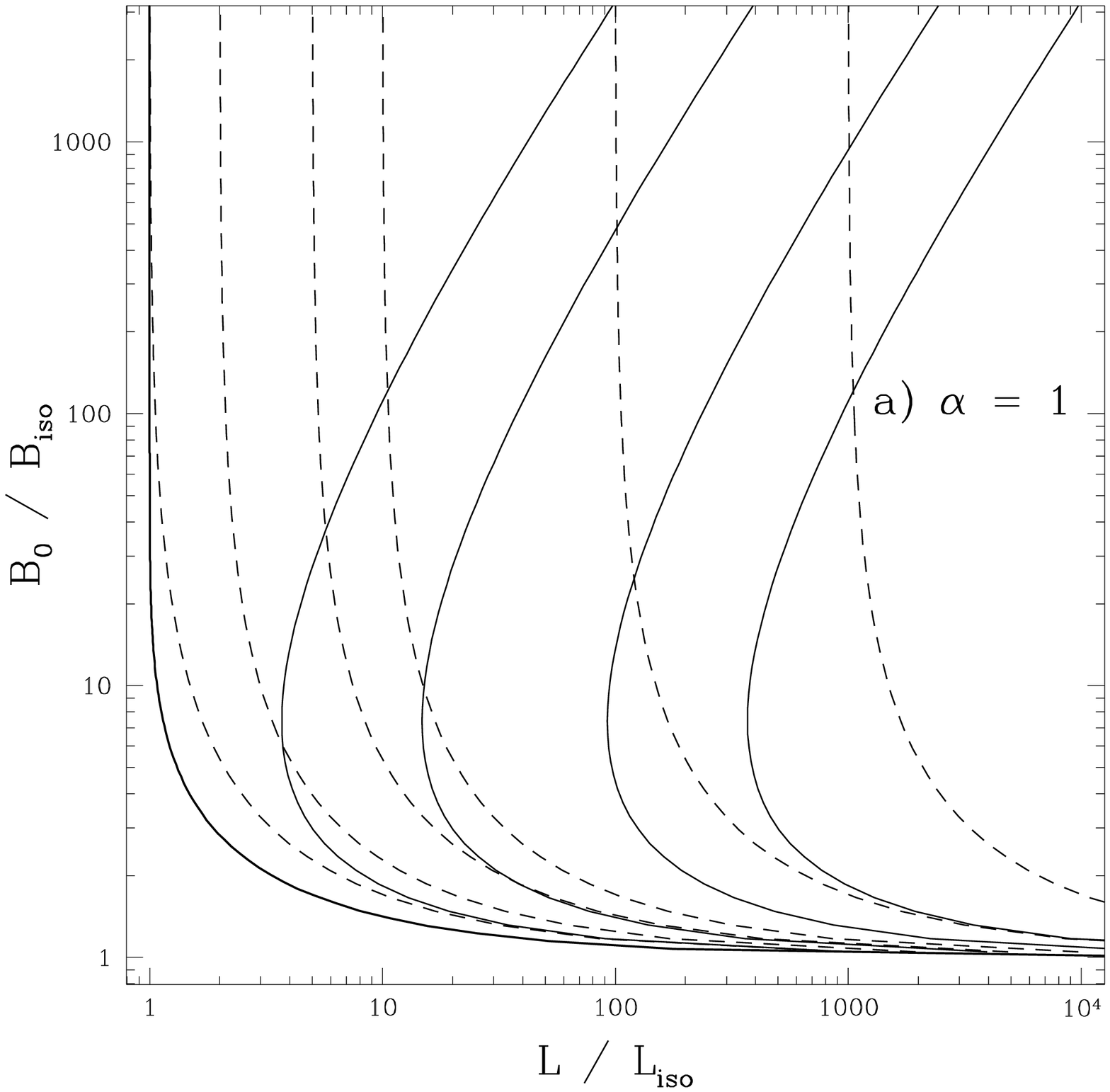,width=0.5\textwidth,height=0.5\textwidth}
\psfig{file=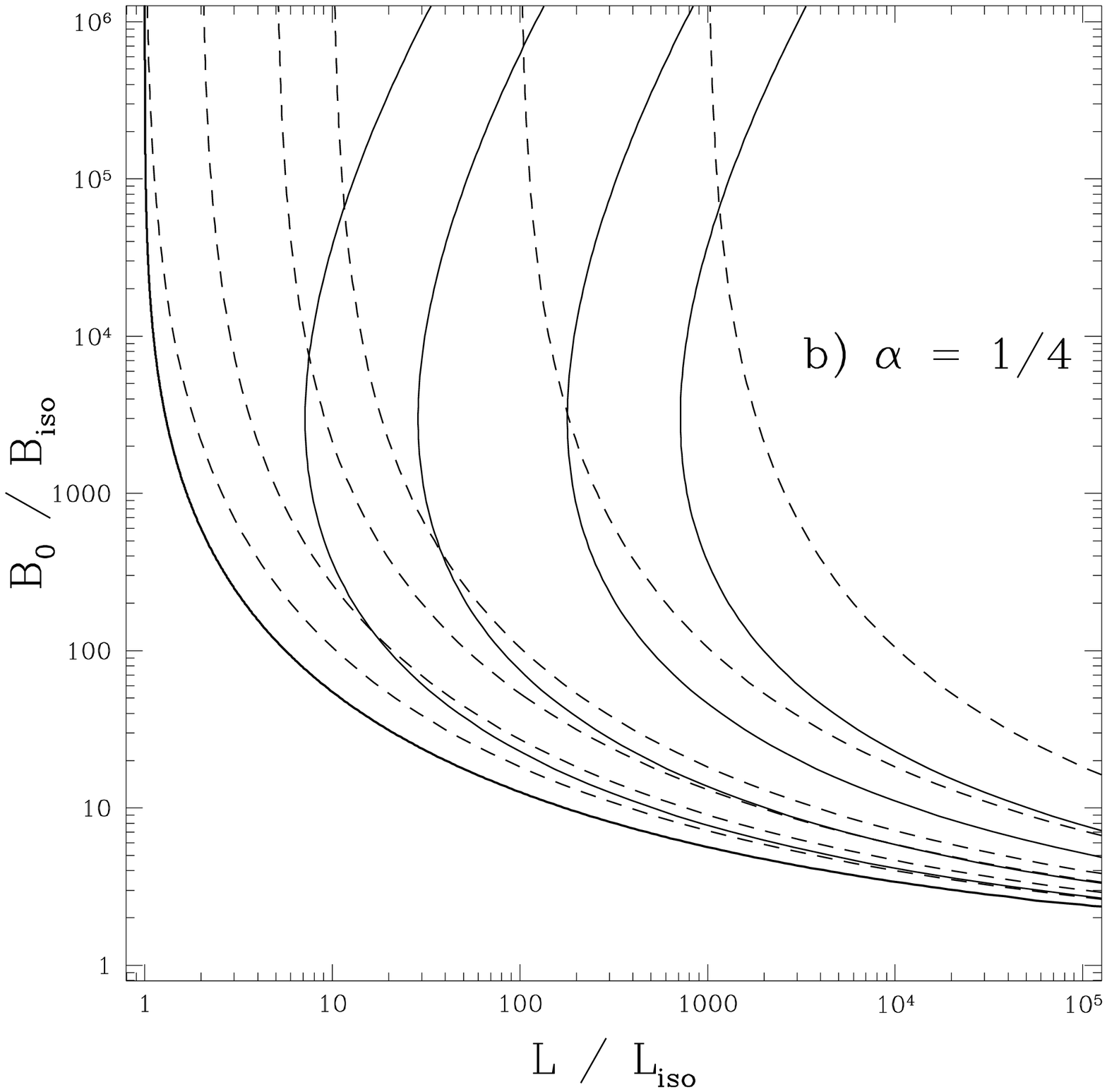,width=0.5\textwidth,height=0.5\textwidth}
}
\caption{
Same as Fig. \ref{B0r0} except for the $B_0-L$ plane. 
The curves are plotted for the
same limiting values as in Fig. \ref{B0r0}.  The luminosity unit $\Liso
=4\pi d_L^2(\pi \t_s^2\biso)$. 
}
\label{B0L}
\end{figure}

It is clear, therefore, that the relation for the counts of extended sources
is considerably more complicated than the commonly used relation 
(\ref{Npoint}).
In order to see these differences more clearly we can rewrite the 
above expressions in terms of the luminosity $L$. For example, if we replace 
$r_0$ by $L=(4\pi )\pi r_0^2B_0F(\infty)$
we can rewrite all the boundary 
conditions in terms of $B_0$ and $L$ instead of $B_0$ and $r_0$.
The three limits in equations (\ref{b1}),
(\ref{b2}) and (\ref{b3})
now give, respectively,  the conditions
\beq\label{b1'}
	B_0 \geq \Biso\left(F^{-1}(\lambda)\right)^2/\lambda, \ \ 
	{\rm with} \ \  \lambda=4\pi d_L^2(\pi \t_s^2 \biso)F(\infty)/L,
\eeq
\beq\label{b2'}
	L \geq L_{{\rm min},\t} \equiv 4\pi d_L^2(\pi \t^2 \biso)
	F(\infty)/\left(\xiso^2f(\xiso)\right),
\eeq
and
\beq\label{b3'}
	L \geq L_{{\rm min},l} \equiv \Lm F(\infty)/F(\xiso).
\eeq
Here $F^{-1}$ is the inverse function of the luminosity curve of growth $F$
and $\xiso$ is a function of $B_0/\Biso$ (eq. [\ref{b3}]).
Figure \ref{B0L} shows the truncations produced by the above
selection criteria in the $B_0-L$ plane.

If we define the distribution $\bar \psi(B_0, L, z)=
\psi(B_0, L, z) dr_0/dL$,
equation (\ref{Next}) 
then becomes
\beq\label{Next'}
	\Nltz={dV\over dz} \left(\int^{B_{0,cr}}_{\Biso} dB_0
        \int^\infty_{L_{{\rm min},l}} dL \bar \psi (B_0,L,z) +
        \int^\infty_{B_{0,cr}} dB_0
        \int^\infty_{L_{{\rm min},\t}} dL \bar \psi (B_0,L,z) \right).
\eeq 

\begin{figure}[htbp]
\leavevmode\centering
\psfig{file=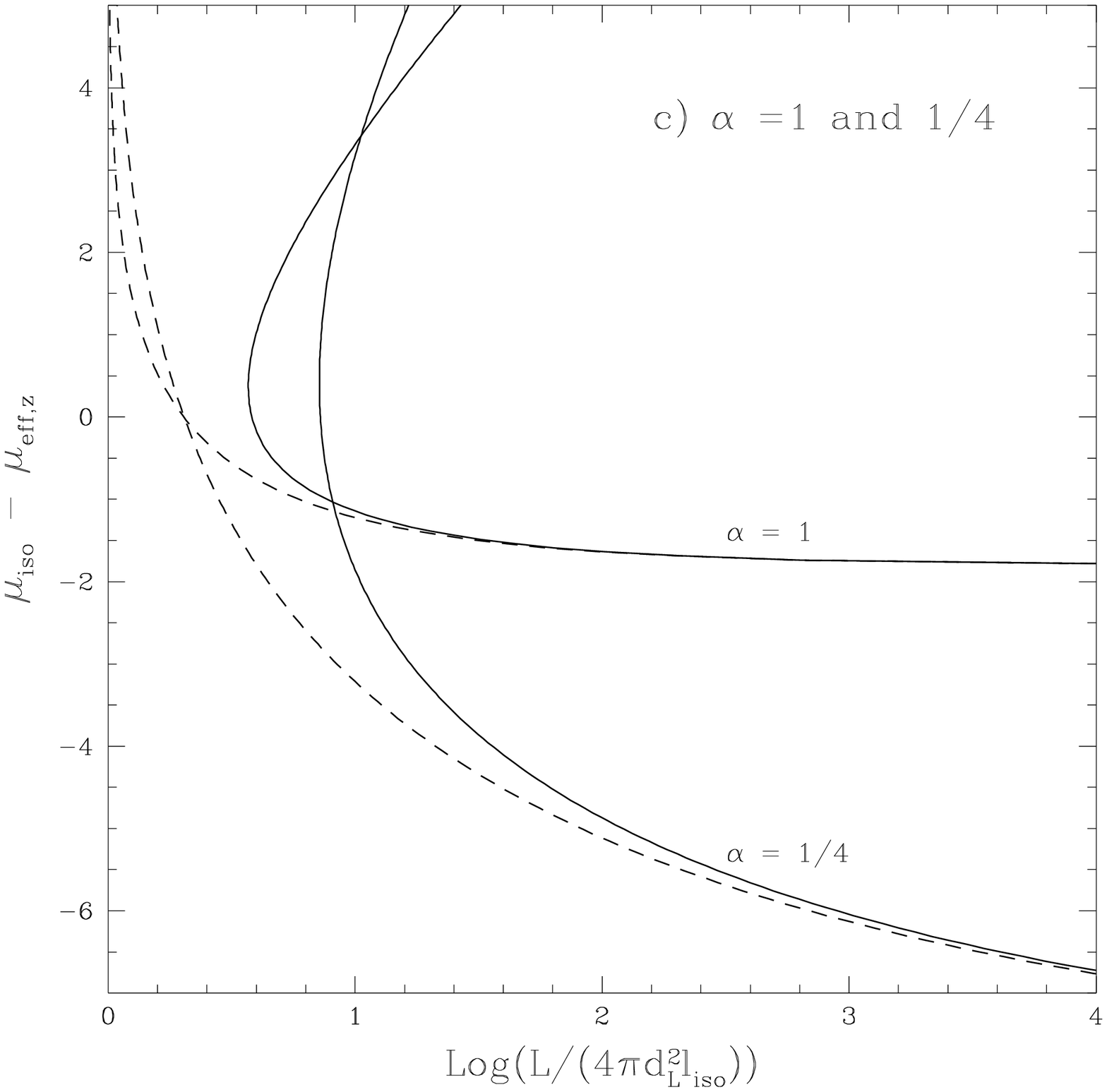,width=0.8\textwidth,height=0.6\textwidth}
\caption{
Same as Fig. \ref{B0L} except we plot the combined size and
flux limits in terms of the {\em effective} values of the
\sb\ in magnitudes; $\mu_{\rm iso}=-2.5{\rm Log}(\biso)+$\ const. and
$\mu_{{\rm eff},z}=-2.5{\rm Log}(B_{\rm eff})+10{\rm Log}(1+z)+$\ const.. 
For clarity, only curves for 
$\t = \t_s$ and $l = \pi \t_s^2\biso$ are shown.
}
\label{B0LL}
\end{figure}

So far we have expressed our results in terms of the central \sb\ $B_0$ 
and core radius $r_0$. As mentioned in \S 1 these parameters are not 
convenient for compring the results with observations. This task can be 
carried out more readily if we express the above relations in terms of 
observationally more meaningful parameters such as the effective
radius and \sb\ defined in \S 1. In Figure \ref{B0LL}
we show the size and flux limits in the \sb-luminosity plane 
for both profiles, where instead of the surface brightnesses $\biso$ and $B_0$
we use the more familiar magnitudes;
$\mu_{\rm iso} = -2.5 {\rm Log}\biso + $\ const., and $\mu_{{\rm eff},z} 
= -2.5{\rm Log}B_{\rm eff} + 10 {\rm Log}(1+z) + $\ const.. 
This figure demonstrates that the 
region of $B_0-L$ plane accessible to observations
is different for the two profiles and shrinks with increasing redshift.

\subsubsection{Comparison With The Point Source Approximation}	

There are several ways that this correct description differs from the
approximate expression given by \eq\ (\ref{Npoint}). Some of these
were discussed by Yoshii (1993).
The first difference is the existence of the second set of integrals
in \eqs\ (\ref{Next}) and (\ref{Next'}) which we discussed above.
Even at high values of the flux limit, $l \gg \lt$, when this additional 
term is negligible  there are two other important differences. The first
is due to the presence of the ratio $F(\infty)/F(\xiso)$ in the lower
limit of the luminosity $L_{{\rm min},l}$, which is absent from 
the lower limit in \eq\ (\ref{Npoint}). 
The second effect is due to the breadth of the distribution of $B_0$. 
For narrower distributions this effect is smaller and disappears 
for a delta function distribution of $B_0$.

These differences can be seen in Figures 4a and 4b as follows. For a given
value of $l$ or the ratio $l/\pi \t_s^2\biso$ 
the point source approximation given by 
equation (\ref{Npoint}) truncates the $B_0-L$ plane by
a vertical line
at the asympthote of  dashed line appropriate for this ratio. 
Then counts all the
sources to the right of this line ($L \geq \Lm$), irrespective of their
surface brightness $B_0$, and uses 
$\phi(L, z)=\int_0^\infty{\bar \psi} (B_0, L, z) dB_0$.
Equation (\ref{Next'}), on the other hand, indicates that this is 
an overestimation of the counts and that we should count only sources
which lie to the right of both a solid and a dashed lines
appropriate for the ratios $\t/\t_s$ and $l/\pi \t_s^2\biso$, respectively. 
It is, therefore, clear that
ignoring the \sb\ and size limitations can cause significant errors 
the extent of which can be quantified only if we know the 
distribution function $\bar \psi$.
We have a good knowledge of the dependence of $\bar \psi$ on $L$ at low
$z$ and high $B_0$ but we have only scanty information on its form at high 
$z$ and low values of $B_0$. The primary aim of the cosmological
tests under the discussion here is to determine the variation with redshift
of the general distribution function $\bar \psi$. A detailed investigation
of these aspects are beyond the scope of this paper. Here we make some 
simple comparisons
between the extended and point source results which do not require
a knowledge of the distribution of $B_0$.

\begin{figure}[htbp]
\leavevmode
\centerline{
\psfig{file=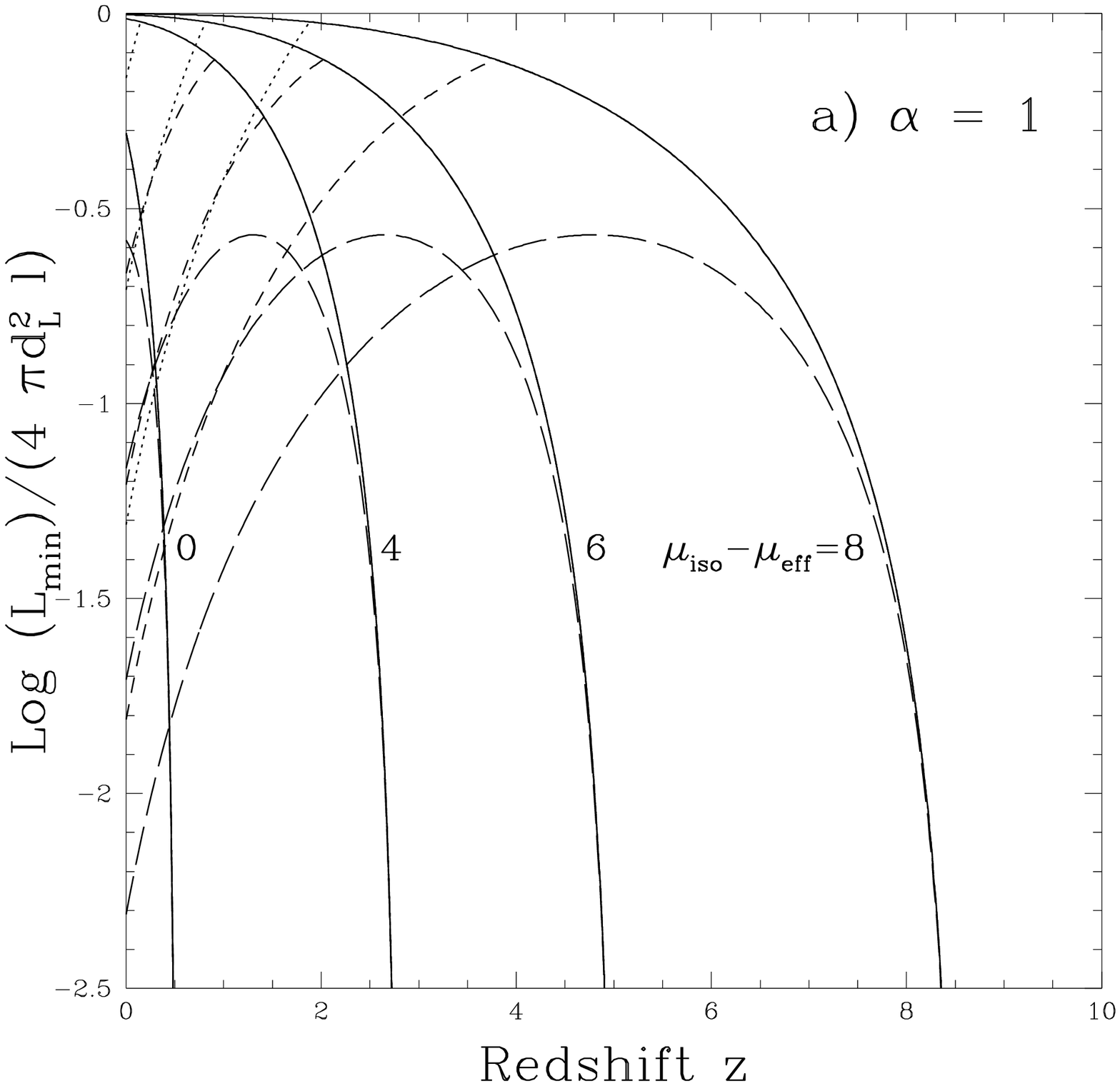,width=0.5\textwidth,height=0.5\textwidth}
\psfig{file=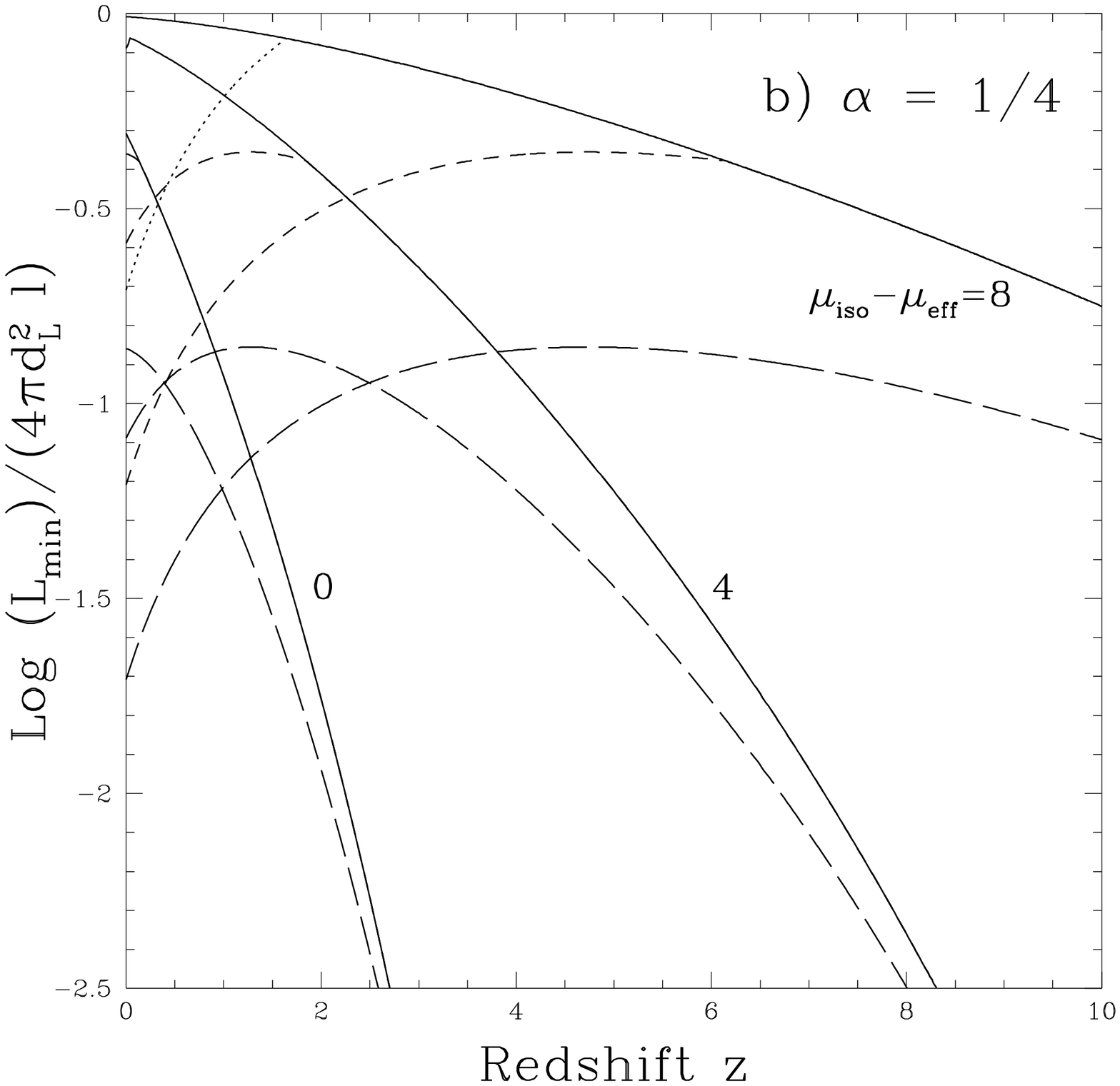,width=0.5\textwidth,height=0.5\textwidth}
}
\caption{
The ratio of the minimum detectable luminosity for extended sources in
the isophotal scheme [eqs. (\ref{b2'}) and (\ref{b3'})]
to that of point sources versus redshift, for indicated values of the
effective surface brightnesses (more exactly the ratio $B_0/\biso$ )
and for four values of the ratio 
$\varrho = l/(\pi \t^2\biso)=1, 3.16, 10 \ {\rm and}\ \infty$, represented
by long dashed, short dashed, dotted and solid lines, respectively.
a)for an exponential (disk) profile, ${\rm ln}f=-x$;
b) for a de Vaucouleurs profile ${\rm ln}f=-x^{1/4}$.
Note the rapid suppression of the flux, as in Fig. 2, when we approach 
the maximum redshift $z_{\rm max}=(B_0/\biso)^{1/4}-1$,
where the relation of $B_0$ and $\biso$ with  $\mu_{\rm eff}$ 
and $\mu_{\rm iso}$, respectively,
are described in the text. Note also the differences between disks 
and spheroids as in Fig. \ref{FandAvsz}.
}
\label{Lmins}
\end{figure}

1){\em Luminosity limits}. The ratio of the limiting luminosities
$\Lminl$ and/or $\Lmint$ to the limit $\Lm$ of the point source approximation,
which depends only on the surface profile, are shown in  Figure 
\ref{Lmins}, for $\alpha=1$ and 1/4, respectively, 
and for several values of
the \sb\ (actually the ratio $B_0/\Biso$) and the 
ratio $\varrho = l/(\lt)$. We use the effective rather the central
\sb\ and express the above ratios in magnitudes. 
As evident for high values of the  ratio $\varrho$, 
i.e. for higher flux limits,
there is an increasing bias against detection of extended sources
at higher redshifts. At lower values of this ratio there is additional
bias against detection of galaxies at low redshifts due to the size limit.

\begin{figure}[htbp]
\leavevmode
\centerline{
\psfig{file=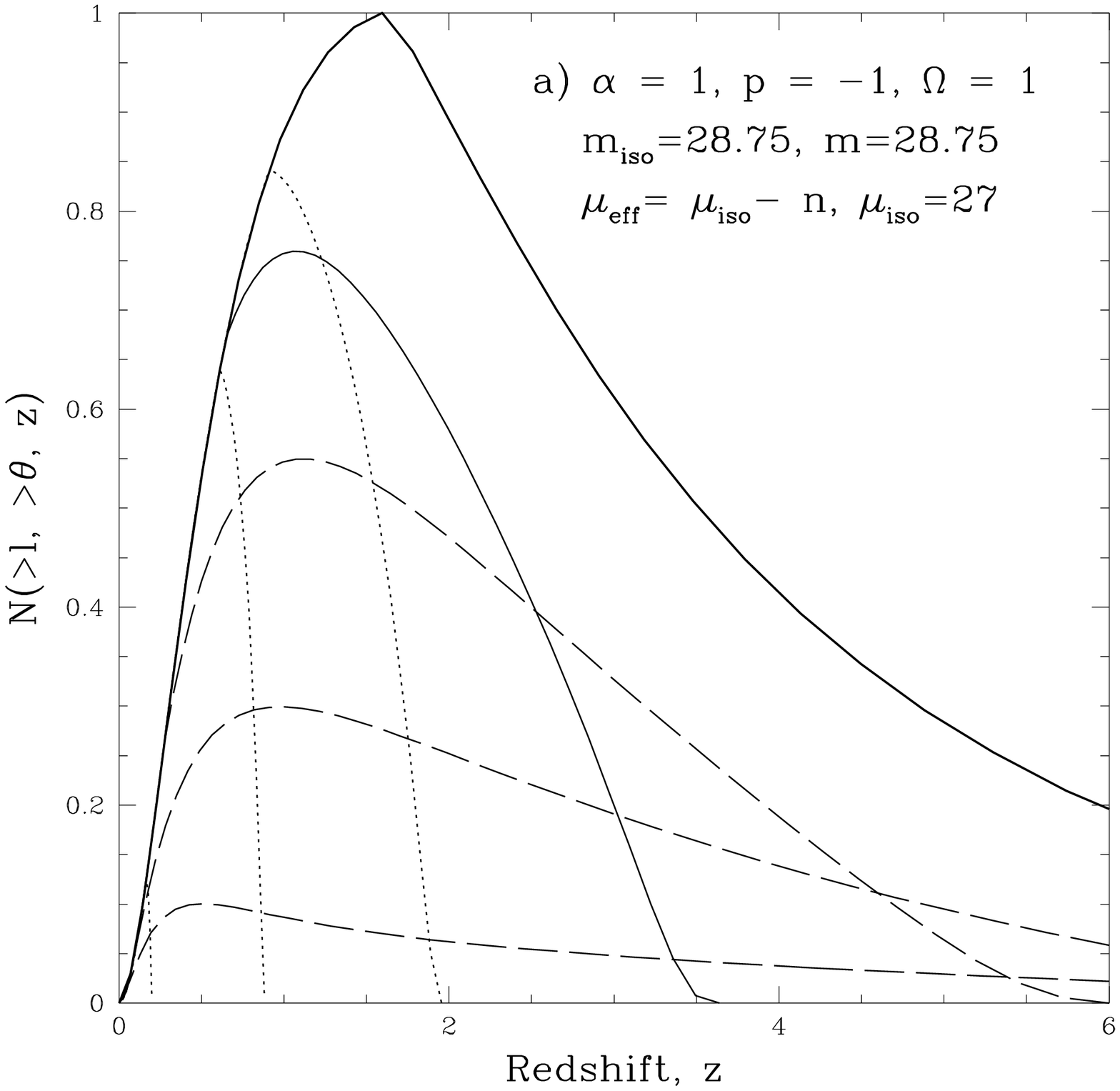,width=0.5\textwidth,height=0.4\textwidth}
\psfig{file=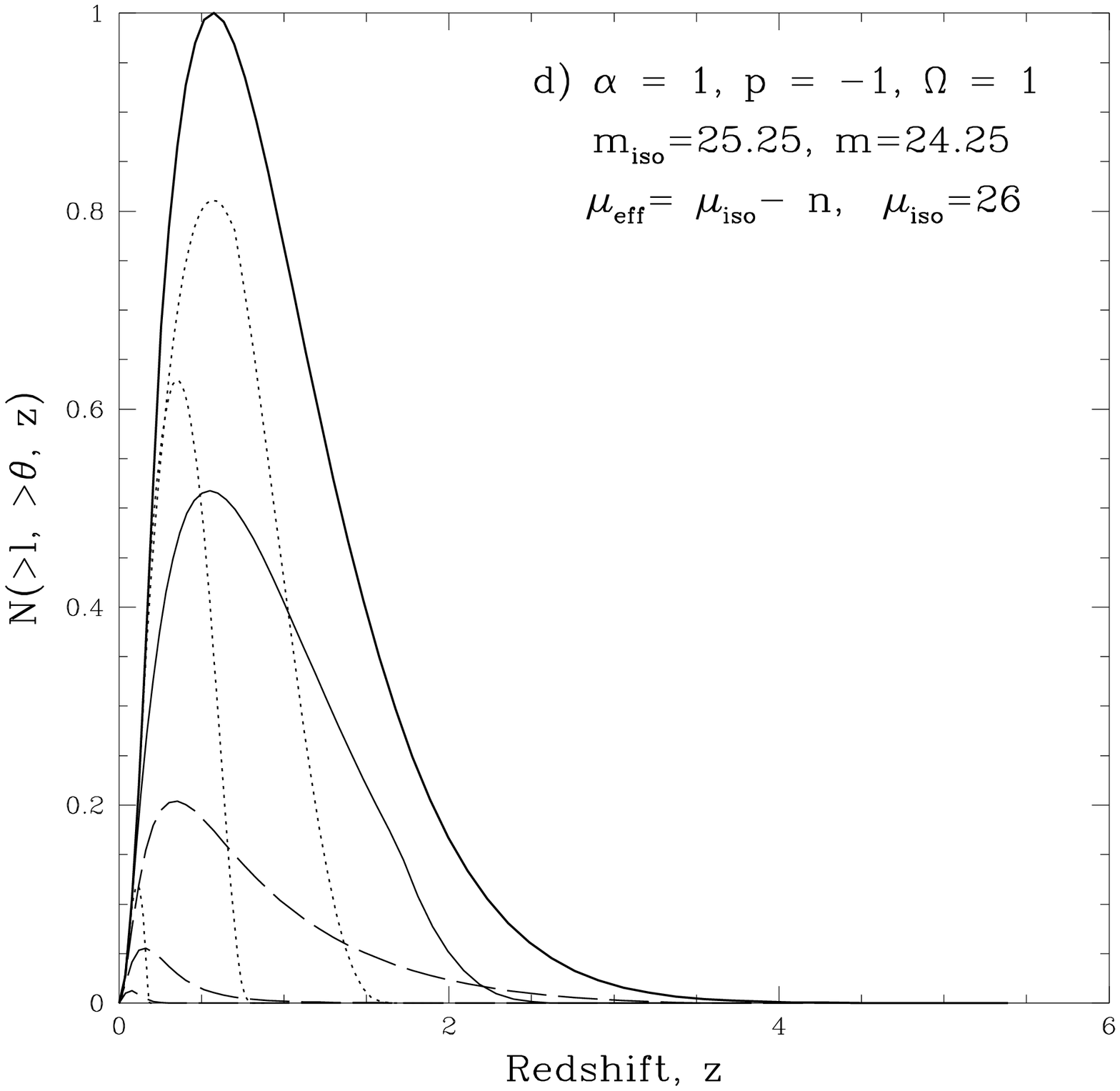,width=0.5\textwidth,height=0.4\textwidth}
}
\centerline{
\psfig{file=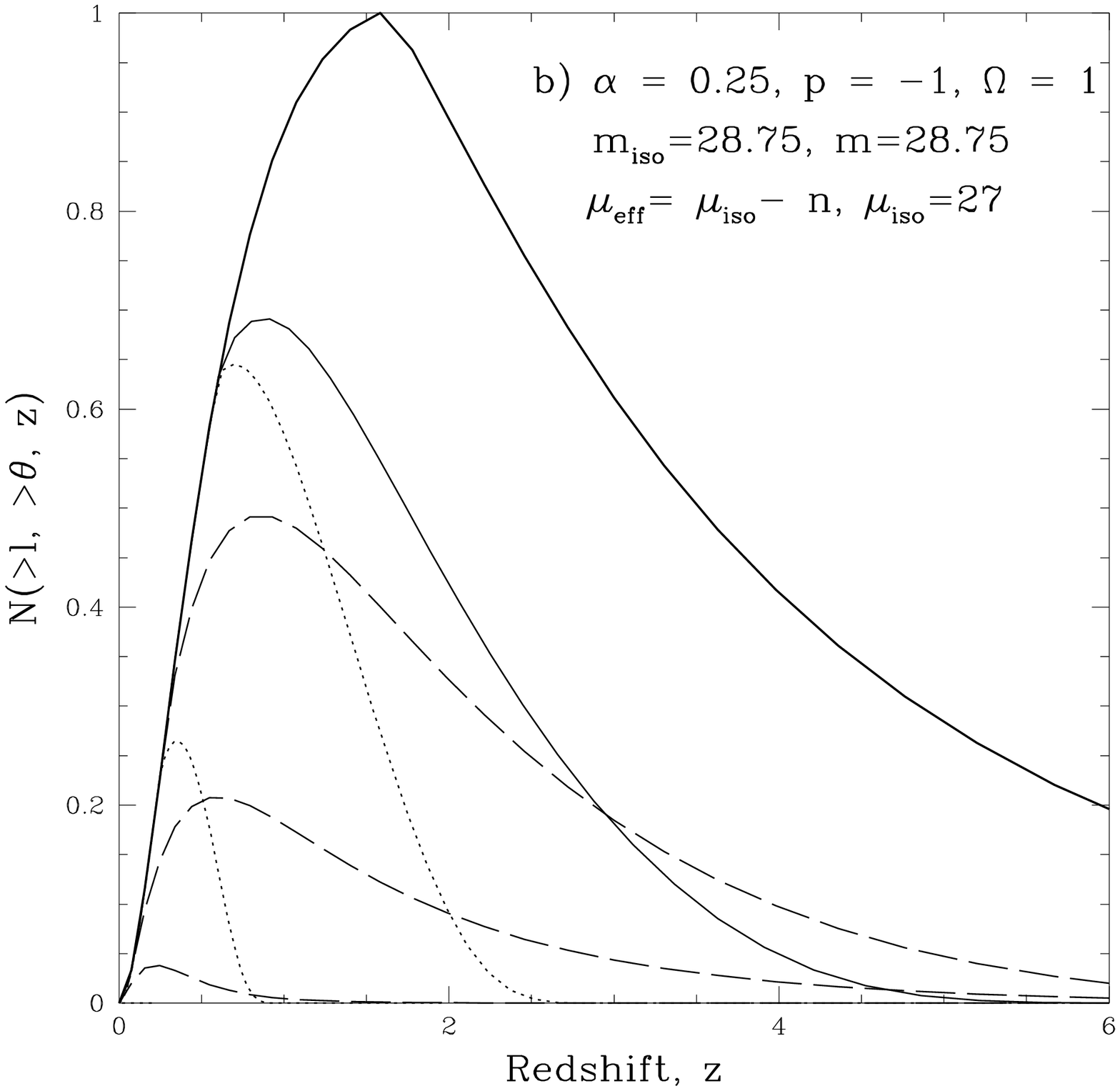,width=0.5\textwidth,height=0.4\textwidth}
\psfig{file=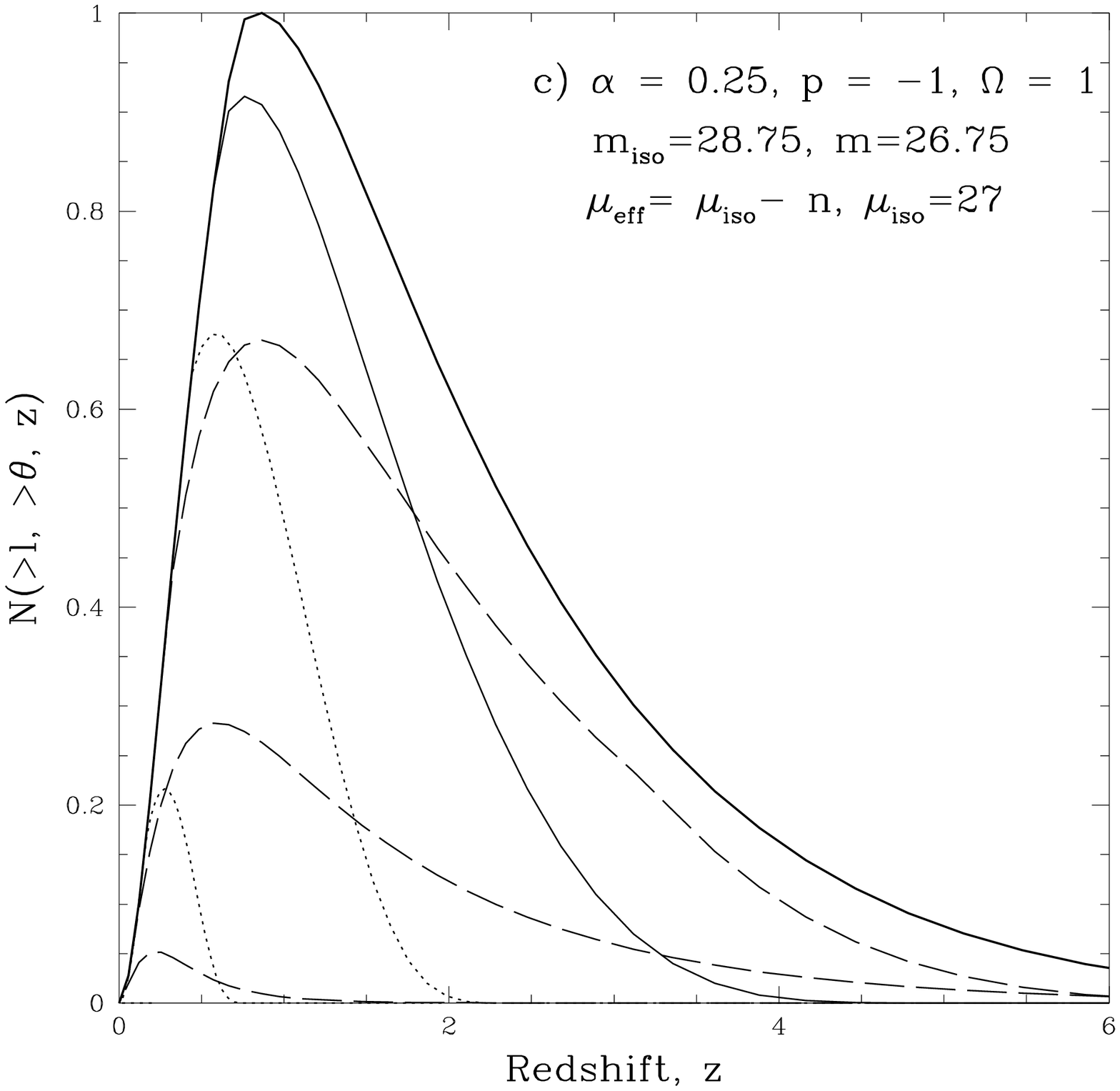,width=0.5\textwidth,height=0.4\textwidth}
}
\caption{
The redshift distribution $N(>l,z)$ of equation(\ref{Np})
of the point sources (top heavy solid lines) and 
extended sources, equation (\ref{Nes}), for 
seven different values of the effective surface brightness.
All curves are normalized by the peak value of the point source distribution.
In each case the dotted lines are for
low surface brightnesses ($n = -1$, 1 and 3) falling below the turning point 
of the solid lines in Fig. 4c, where the \sb\ limit is most important,
and the dashed lines are for high surface brightnesses ($n = 7$, 9 
and 11) above this turning point, where the size limit is important. 
The thin solid line (with $n = 5$) lies generally near the turning point.
The limiting magnitude is 
$m = 2.5{\rm log}l\  + \ $const  and $m_{\rm iso} = -2.5{\rm log}
(\pi \t_s^2 \biso)\ + \ $const.. The \sb\ 
is in units of magnitude per square arcsec.
{\em a} and {\em d} for disks,  and
{\em b} and {\em c} for spheroids  with different limits.
}
\label{Nofz}
\end{figure}

2){\em Redshift Distributions}. These differences can also be seen
when we compare the redshift distributions
expected for point and extended sources. 
For the purpose of illustration let us assume that $L$ and $B_0$
are uncorrelated (clearly not a good assumption)
so that we can separate the distribution function  as 
$\bar \psi =\rho (z) h(B_0) \phi (L/L^*)/L^*$, where $\rho (z)$ and $L^* (z)$
describe the density and luminosity evolution of the sources, and $h(B_0)$
gives the distributions of the central surface brightness.
If we define the cumulative functions $H(B_0)=\int_0^\infty h(B_0)dB_0$ and
$\Phi(x)=\int_x^\infty \phi (x')dx'$, with $H(0)=\Phi (0)=1$, then 
equations (\ref{Npoint}) and (\ref{Next'}) become, respectively
\beq\label{Np}
	\Nlz={dV \over dz} \rho (z)\Phi (\Lm/L^*)
\eeq
and
\beq\label{Ne}
	\Nltz={dV \over dz} \rho (z)\left(\int_{\Biso}^{B_{0,cr}}
	h(B_0)\Phi (\Lminl/L^*)dB_0 + \int_{B_{0,cr}}^\infty
	h(B_0)\Phi (\Lmint/L^*)dB_0\right).
\eeq
In absence of an exact knowledge of the distribution $h(B_0)$,
we compare these expressions for different assumed values of the central 
or effective \sb\, which amounts to a delta function approximation 
of $h(B_0)$. In this case \eq\ (\ref{Ne}) simplifies to
\beq\label{Nes}
        \Nltz={dV \over dz} \rho (z)\cases{\Phi (\Lminl/L^*), &if 
        $B_{0,cr}>B_0$;\cr
        \Phi (\Lmint/L^*), &if  $B_{0,cr}<B_0$.\cr}
\eeq
Assuming a Schechter luminosity function, $\phi(x)\propto x^p e^{-x}$,
we evaluate the redshift distributions
for some representative values of the \sb\ $B_0$ (or $\mu_{\rm eff}$)
and for several combinations of the limits $l$ (or magnitude
$m$), $\t$ and $\biso$ (or $\mu_{\rm iso}$).
The results are shown in Figure \ref{Nofz} 
for $p= -1$, and 
for a cosmological model with $\Omega =1$ and $\Lambda = 0$.
We also assume absence of any density or luminosity evolutions; 
$\rho$ and $L^*$ constants.

It should be noted that the relative shapes of the point source and 
various extended source distributions are independent of the 
cosmological model or the density evolution $\rho(z)$. As evident
equation (\ref{Npoint}) or (\ref{Np}) give quite incorrect redshift 
dependences, overestimating the number of sources by a large factor 
at high redshifts,
specially for low values of the \sb\ due to the
\sb\ limit, and at high values of \sb\ due to the size limit. Clearly 
ignoring these effects could lead to incorrect results. For example,
if these expressions were used to derive the
extent of the luminosity evolution, $L^*(z)$, they would underestimate
this evolution by  factors equal to $F(\xiso)/F(\infty)$ and 
$\xiso^2f(\xiso)/F(\infty)$, depending whether 
$l/(\lt) \gg 1$ or is equal to 1, respectively.
The situation is more complicated when the effects of
the dispersion of the \sb\ or 
its correlation with the luminosity (or core size $r_0$) are included. 
In such cases there would be errors in the determination of the 
density evolution as well.

\begin{figure}[htbp]
\leavevmode
\centerline{
\psfig{file=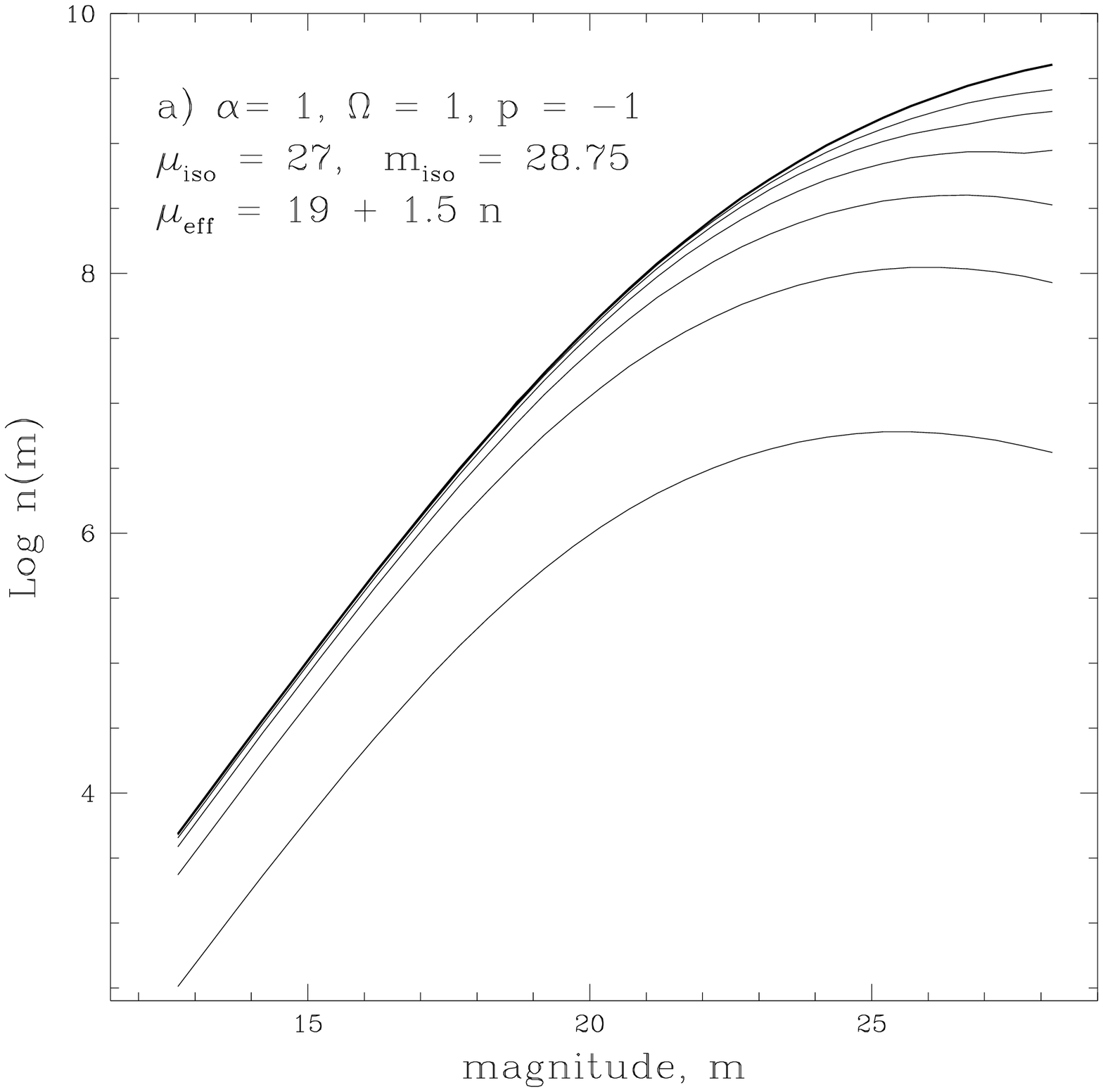,width=0.5\textwidth,height=0.5\textwidth}
\psfig{file=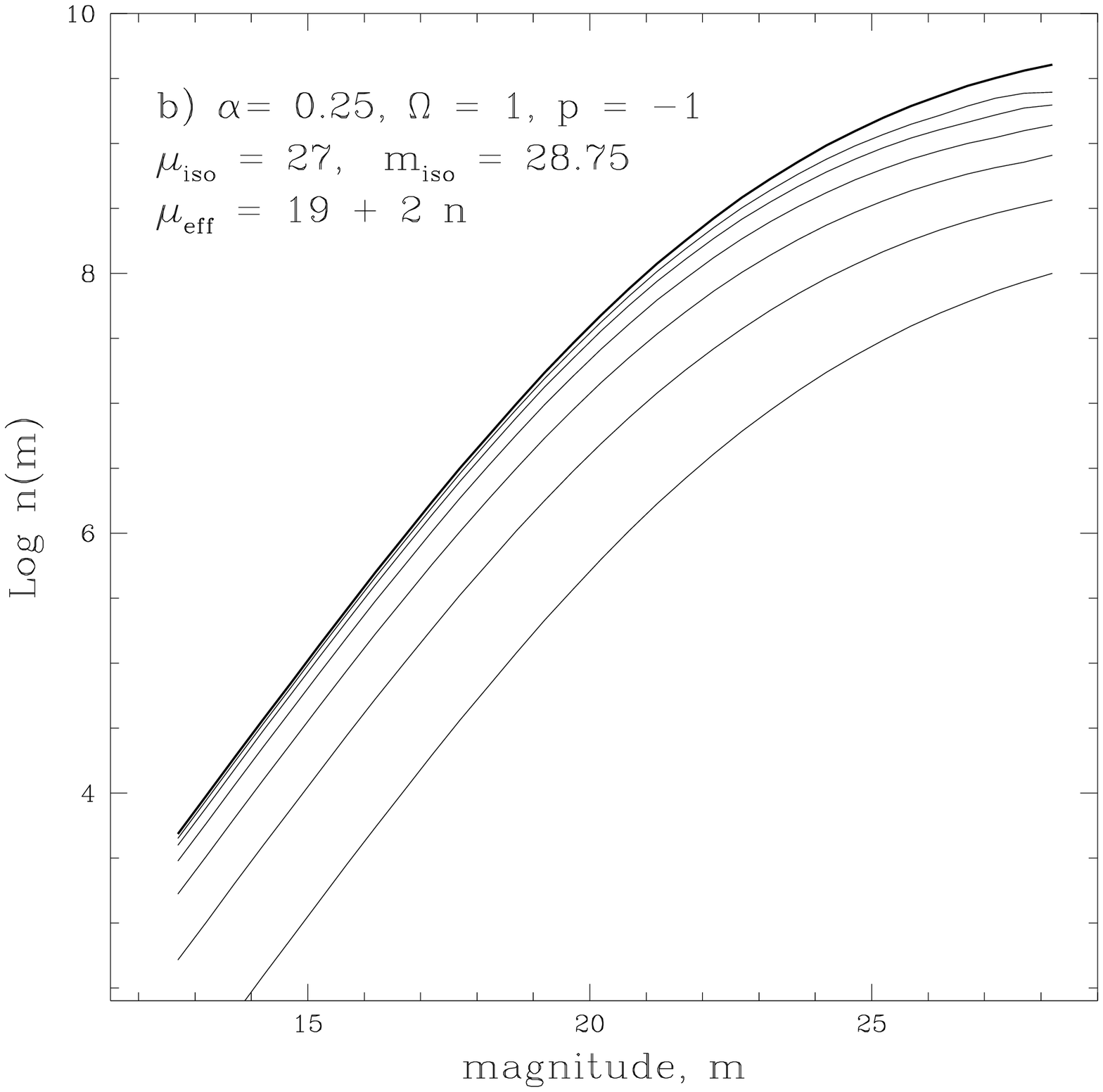,width=0.5\textwidth,height=0.5\textwidth}
}
\caption{
The differential counts of point sources (top thick line)
and that of extended sources  with different \sb\
limits (thin lines with $n = 1$ to 6 from top to bottom)
for the same model parameters as Fig. \ref{Nofz}.  
a)for an exponential (disk) profile, ${\rm ln}f=-x$;
b) for a de Vaucouleurs profile ${\rm ln}f=-x^{1/4}$.
Note the reduction at large magnitudes for high \sb\
sources and large suppression at all magnitudes for low \sb\ sources.
}
\label{diffc}
\end{figure}

3){\em Source Counts.} Integrating the redshift distributions over $z$
gives the cumulative counts. Differentiation of this gives the 
differential counts. For example, for point sources $n(l) =
\int_0^\infty {dV \over dz}\phi(\Lm, z)(4\pi d_L^2)dz$.
Figure \ref{diffc} shows the differential magnitude counts 
of extended sources with various values of \sb\, as well as 
that of point sources.
We use the same model parameters as above. 
Again, as evident
the neglect of the selection aspect discussed above can cause
considerable error in the determination of the evolution of the 
general luminosity function or the cosmological parameters.

\subsection{Point Sources Revisited}

It should be noted that some of these effects are present even for
unresolved or point sources. However, in this case the correct 
equation is only slightly different than  \eq\ (\ref{Npoint}) or (\ref{Np}).
When $\t_s d_A \gg r_{\rm eff}$ the sources are unresolved and they all
have essentially the same profile $g$ and size $\t_s$  as the PSF;
$\hat B(\t) = B_0\xi(\t_s d_A/r_0)g(\t/\t_s)$. The total luminosity 
can be written as $L= 4\pi d_A^2(\pi \t_s^2)B_0\xi$. 
If we limit ourselves to the isophotal fluxes $>l$ and sizes $>\t$, then
the three selection criteria (eqs. [\ref{b1}], [\ref{b2}] and [\ref{b3}]) 
become almost identical in form;
$B_0 \geq \Biso/\left(K_i\xi(\t_s d_A/r_0)\right)$,
where $\xi$ is defined in \eq\ (\ref{psfB}) and $K_i =1, g(\t/\t_s)
\ {\rm and}\ (4\pi)(\pi \t_s^2\biso)G(\tiso/\t_s)/l$, respectively for 
the three limits. In the above relations we have defined the
cumulative PSF as 
$G(x)=\int^x_0 2g(x)xdx$, $G(\infty)=1$, and $\tiso$ is obtained from 
$g(\tiso/\t_s)=\Biso/(B_0\xi) = \Liso /L$. 
Clearly the size limit does not make sense
for unresolved sources and the flux limit
is the most restrictive limit. It can be shown then that the correct
expression for the source counts is 
\beq\label{Npoint'}
        \Nlz={dV\over dz} \int^\infty_{\Biso} dB_0
        \int^\infty_{L_{\rm min}, s} dL \bar \psi (B_0,L,z),
\eeq
where the lower limit of the luminosity is obtained from the solution of 
\beq\label{Lmp}
	L G\left(g^{-1}(\Liso/L)\right)=4\pi d_L^2 l, \ \ 
	{\rm with} \ \ \Liso\equiv 4\pi d_L^2 (\pi \t_s^2 \biso).
\eeq
Here $g^{-1}$ is the inverse function of the PSF.
In addition to the \sb\ cutoff and the integration over $B_0$, this expression
differs from the simple \eq\ (\ref{Npoint}) also by the 
presence of the term involving the cumulative PSF $G$ in the
integration limit. This difference becomes important only 
for flux limits very near the isophotal values; $l=\pi \t_s^2 \biso$.
The \sb\ limit can be important for unresolved galaxies because of 
their low intrinsic \sb\ or effective temperatures. But for 
other point sources such as quasars whose \sb\ is equal to that of a hot
accretion disk this effect is negligible (becoming important
only at extremely high redshifts) and the point source
approximation of \eq\ (\ref{Npoint}) is very accurate.

\subsection{Combined Counts}

In principle we can combine the counts of the resolved and unresolved sources
by replacement of the $B(r)$ with the modified profile $\hat B(r)$
of \eq\ (\ref{psfpr}).  
We can then repeat the procedure carried out for the extended sources
with the replacement of the profiles $f$ and $F$ with $\hat f$ and $\hat F$
and change the limits correspondingly, except now the profiles are 
functions of the additional parameter $\t_s d_A/ r_0$.
However, now the size limit
is unnecessary because we can include all sources. Of course one 
must make sure that the sample of galaxies, for example,
is not contaminated by other unresolved sources (e. g. stars).
We therefore have the simpler expression
\beq\label{Nall}
        \Nlz={dV\over dz} \int^\infty_{\Biso} dB_0
        \int^\infty_{\hat L_{\rm min}, l} dL \bar \psi (B_0,L,z),
\eeq
where $\hat L_{{\rm min}, l}$ is given by \eq\ (\ref{b3'}) with 
$F(\xiso)\rightarrow \hat F(\xiso)$; note that 
$\hat F(\infty) = F(\infty)$. The parameter  $\xiso$ now is obtained 
from $\hat f (\xiso, \t_s d_A/r_0)=\Biso/\hat B_0$ with
$\pi r_0^2 = L/\left(4\pi B_0F(\infty)\right).$

\subsection{Other Tests}

Differentiation of $\Nltz$ gives the differential distribution $n(l,\t,z)$
from which we can calculate various moments and compare them to observations.
For example, the flux-redshift relation can be obtained from
\beq\label{zm}
        <l(z)>=\int^\infty_l \left(LF(\xiso)/\left(4\pi d_L^2
	F(\infty)\right)\right) \nltz dl\bigg/\Nltz.
\eeq
In a similar fashion one can derive $<\t >-z$ or $<\t >-<l>$ relations. 

\section{CLASSICAL TESTS: METRIC VALUES}

Some of the complications evident in the above analysis can be 
avoided if instead of the isophotal sizes and fluxes we deal
with some {\underbar {metric}} values of these quantities.
For example, if we define a proper metric size $r_p$ corresponding to
a constant value of the function $\eta$, say $\eta_0$, as defined in P76 and
\eq\ (\ref{eta}), then the expressions for the surface brightness,
size and flux limits (or truncations) become considerably simplified.
For a limiting \sb\ $\biso$, a limiting angular radius $\t$ and
a limiting flux $l$ these truncations are described by the following:
\beq\label{bm}
	B_0 \geq \Biso/f(\zeta),  \ \  r_0 \geq d_A\t /\zeta, 
	\ \ {\rm and} \ \ L \geq\Lm F(\infty)/F(\zeta),
\eeq
where $\eta(\zeta)=\eta_0$ and  $\zeta=r_p/r_0$.

Note that in contrast to the complicated truncations we found
for the isophotal case, \eqs\ (\ref{b1}), (\ref{b2}) and (\ref{b3}),
the current truncations are much simpler; they depend only on one 
observational limit and the redshift.
The above limits are good for $\t \gg \t_s$, the size  of the PSF.
This will always be true in this case  because of the need to have 
a well defined \sb\ distribution.

If the truncation
limits are chosen so that $l = \pi \t^2 \biso$, then the 
last limit in equation (\ref{bm}) due to the flux 
limitation falls below the other two and can be ignored. In this case the
data truncation in the $B_0-r_0$ plane is parallel to the axis
making the calculation of the observables straight forward and
free of the complex limits of integration. Since 
the flux limit $l$ does not
enter in the determination of the 
observed distribution of the sources, such a sample will not be  
appropriate data for tests based on source counts as a function of
the flux $l$. But such a sample can be used to obtain the 
distributions of the angular size,
average surface brightness or redshift. For example the latter 
is simply
\beq\label{Nmz}
	N(\zeta, z)={dV\over dz}\int_{\Biso/f(\zeta)} ^\infty dB_0
	\int_{\t d_A/\zeta} ^\infty dr_0 \psi(B_0,r_0,z).
\eeq
Because of this simplification, such samples 
are well suited for tests based on the moments of the observed distributions.
For example the angular size-redshift relation is simply given as
\beq\label{tmz}
	<\pi \t^2(z)>= {\zeta \over {N(\zeta, z) d^2_A}}{dV\over dz}
	\int_{\Biso/f(\zeta)} ^\infty dB_0
	\int_{\t d_A/\zeta} ^\infty dr_0 \pi r^2_0 \psi(B_0,r_0,z).
\eeq
Similarly, for the flux-redshift relation we have
\beq\label{zmm}
	<l(z)>= {F(\zeta) \over {N(\zeta, z) d_L^2}}{dV\over dz}
	\int_{\Biso/f(\zeta)} ^\infty dB_0 B_0
	\int_{\t d_A/\zeta} ^\infty dr_0 \pi  r_0^2 \psi(B_0,r_0,z),
\eeq
or, in terms of the luminosity $L$ 
\beq\label{zmm'}
	<l(z)>= {F(\zeta)/F(\infty) \over {N(\zeta, z) d_L^2}}
	{dV\over dz}
	\int_{\Biso/f(\zeta)} ^\infty dB_0 
	\int_{L_{{\rm min},\zeta}} ^\infty dL L \bar \psi(B_0,L,z),
\eeq
with a similar expression for the size-$z$ relation. The integration limit
\beq\label{bm'}
	L_{{\rm min}, \zeta}= 4\pi d_L^2(\pi \t^2 \biso )(F(\infty)/\zeta^2)(B_0/\Biso).
\eeq
These expressions are considerably simpler than the corresponding \eqs\
for the isophotal analysis.

Similar expressions can be derived for other definitions of the metric
quantities. For example, as mentioned in \S 1 in connection with Figures 
1 and 2, instead of $r_p$ one can use the effective
radius $r_{\rm eff}$, within which resides a certain fraction (usually half) 
of the total light. This would amount to a new definition of the
constant $\zeta$ as $F(\zeta)=F(\infty)/2$. This may be a convenient
procedure for nearby galaxies but not at high reshifts, because it 
relies on the knowledge of the total flux, the determination of
which lies at the heart of the
difficulty associated with these tests. The procedure proposed in P76
relies only on the data within a specified radius and not on the data 
from the outer, invisible parts; $\t_p$ is obtained by setting the ratio of
the average to limiting surface brightnesses to a fixed value.

\section{NEW TESTS}

The discussions in the above two sections demonstrate that an accurate analysis
of the extragalactic data for the purpose of cosmological tests 
is complicated and must include all of the
above mentioned considerations. In particular, it is imperative to  
keep in mind 
the multivariate nature of the problem  and to account for the surface
brightness limitation, (eq. [\ref{Biso}]), common in all of the
above expressions. The dispersions in the distributions of $B_0$, $r_0$
or $L$, and the correlations between these can have substantial effect 
on the final results. These effects are more pronounced when dealing
with the isophotal quantities than with the metric ones. However, the
latter procedure must be limited to well resolved sources, while
the former, in principle, could be extended to unresolved sources
if a good knowledge of the PSF is at hand.

This task, however complicated, can be
carried out given the knowledge of the distribution function and the
brightness profile. With sufficient care in the analysis of the
data and in modeling one can determine either the cosmological evolution
of the sources (i.e. the redshift variation of the  distribution $\psi$)
or the cosmological parameters.
Such analyses, which may be simple or appropriate
for data limited to low reshifts   
is not the simplest method to determine the cosmological or 
galactic evolutions. For example, the traditional method of 
identifying sources with
some apparent flux may not be necessary or be the most straight forward way 
of carrying out this task. The complexities described in the previous sections
are the result of the multiple selection criteria needed for 
counting individual galaxies.
 
We now describe two new and much simpler tests which 
combine the good aspects and avoid the complexities of the two methods 
described above, and are much better
suited for the analysis of modern digitized CCD data. 
The essence of these tests is to reduce the selection criteria to one,
namely the surface brightness, and deal with the distribution of  $B_0$. 
In practice this amounts to simply
counting the number of pixels at a given \sb\ (or adding up
their intensities) independent
of which galaxy they belong to. This way one can
avoid the complexities arising from the need to  define the sizes and
fluxes (isophotal or metric) for every galaxy in the field.

\subsection{Sky Covered By Galaxies}

The first of these tests, which is related
to the angular diameter test, involves computation of the fraction 
of the sky which is covered by all galaxies above (cumulative)
or within (differential) a given range of surface brightness, $b$ to $b+db$.
Observational determination of this fraction is simple. 
It is accomplished by counting the number of pixels with a given intensity 
value. The expressions relating this quantity to the cosmological 
models and to the properties of the galaxies are decidedly (somewhat) simpler
than those for the isophotal (metric) treatment.
Let us first consider well resolved sources, namely those with 
angular radii $\geq \t$ which we take to be $\gg \t_s$.
In this case the only data truncation arises from the size limit
which is same as that described by \eqs\ (\ref{tiso}) and (\ref{b2})
with the isophotal quantities replaced by those for an arbitrary value of $b$.
\beq\label{bsky}
	B_z=\Z4 b, \ \ x_b \equiv \t_b d_A/r_0=f^{-1}(B_z/B_0) \ \ 
	{\rm with} \ \ B_0 \geq B_z/f(\t d_A/r_0).
\eeq
Now following the same steps as in the previous sections
the sky fraction covered by all galaxies down to a
given apparent surface brightness $b$ is obtained by adding
the contribution $\pi \t_b^2=\pi (x_b r_0 /d_A)^2$ of each galaxy. 
The result is 
\beq\label{Fsky}
	{\cal F}_{\rm sky}(>b, z)={d\chi \over dz}\int_{B_z} ^\infty  dB_0
	x_b^2 \int_{\t d_A/x_b} ^\infty 
	dr_0 \pi r_0^2 \psi(B_0,r_0,z),
\eeq
where the line element 
\beq\label{line}
	d\chi/dz=(dV/dz)(4\pi d_A^2)^{-1}.
\eeq
Alternatively, in terms of the luminosity distribution
\beq\label{Fsky'}
	{\cal F}_{\rm sky}(>b, z)={d\chi \over dz}\int_{B_z} ^\infty dB_0
	{x_b^2 \over {B_0 F(\infty)}} 
	\int_{L_{{\rm min}, b}} ^\infty dL L \bar \psi (B_0,L,z),
\eeq
where
\beq\label{Lmb}
	L_{{\rm min}, b}= 4\pi d_L^2(\pi \t^2 b)F(\infty)/
	\left(x_b^2f(x_b)\right).
\eeq
Note that these equations have the simplicity of the tests based on
metric values; \eq\ (\ref{Fsky}), except for the term $x^2_b$ is identical to
$(<\pi \t^2(z)>)N(\zeta, z)$, shown in equation (\ref{tmz}). However,
more importantly, the data analysis is enormously simpler because it
does not require determination of the \sb\ profile and the metric or isophotal
values for each galaxy.

As in the case of the classical tests, this test also 
can be carried out for unresolved
sources. But this is not much different than counting sources because
all unresolved sources have essentially the same area, $\pi \t_s^2$.
Following the procedure in \S {\bf 2.3} it can be shown that
the sole truncation due to the \sb\ limit, $B_0\xi \geq B_z$, 
is equivalnt to 
\beq\label{Lb}
         L = 4\pi d_A^2(\pi \t_s^2B_0)\xi \geq L_b \equiv
         4\pi d_L^2(\pi \t_s^2b)
\eeq
and the actual angular radius of each source is given as $\t_b / \t_s
=g^{-1}(B_z/B_0\xi) = g^{-1}(L_b/L)$. Thus, the fraction of the sky
covered by unresolved sources then becomes
\beq\label{Fskyp}
        {\cal F}_{\rm sky}(>b, z)={dV \over dz} (\pi \t_s^2)
	\int_{B_z} ^\infty dB_0
        \int_{L_b} ^\infty dL \left(g^{-1}(L_b/L)\right)^2
	\bar \psi (B_0,L,z).
\eeq

Similarly, as in \S {\bf 2.4}, if we have a good knowledge of 
the form of the PSF we can combine resolved and unresolved sources
by expressing the above relations in terms of the modified profile
$\hat f$ as
the convolution of the profile $f$ and the PSF $g$ [eq. (\ref{psfpr})]. 
Then we do not need not to specify
a size limit and the only truncation comes from the value of
the apparent \sb\ $b$. However, in this case we have the added complication
due to the dependence of the characteristics on the ratio $\t_s d_A/r_0$.
For example, \eq\ (\ref{Fsky'}) now becomes
\beq\label{Fskyall}
	{\cal F}_{\rm sky}(>b, z)={d\chi \over dz}\int_{B_z} ^\infty  
	{{dB_0} \over {B_0F(\infty)}}
	\int_{\hat L_{{\rm min}, b}} ^\infty dLL 
	x_b^2 \bar \psi (B_0,r_0,z),
\eeq
where now $\hat L_{{\rm min}, b}$ is given in equation (\ref{Lmb}) 
with $f \rightarrow {\hat f}, F \rightarrow {\hat F}$ and 
$x_b=\t_b d_A/r_0$ is a function of both $B_0$ and $L$ (or $r_0$) 
and is obtained from the inversion of the relation
$\hat f (x_b, \t_s d_A/r_0)=B_z/B_0\xi$, with $\pi r_0^2=L/(B_0F(\infty))$.

The differential distribution can be obtained from
$f_{\rm sky}(b,z)=-\partial {{\cal F}_{\rm sky}(>b,z)}/\partial b$. The 
integration of either distribution over $z$ gives the differential 
or cumulative distributions of all galaxies irrespective of their redshift.

\subsection{Total Sky Brightness}

The second test, which is related to the flux-redshift test, deals with the 
contribution of all galaxies to the sky brightness within a range of
(or above) a given surface brightness $b$. 
This amounts to adding all of the intensity
values of the appropriate pixels. This is to be then
compared with the expression for the total intensity (flux per sterradian)
as the sum of the contribution $l(\t_b)= \pi r_0^2 B_0 F(x_b)/(4\pi d_L^2)$
of all resolved galaxies with $\t_b \geq \t$. 
\beq\label{Isky}
	{\cal I}_{\rm sky}(>b,z)={1 \over {\Z4}} {d\chi \over dz}
	\int_{B_z} ^\infty dB_0 {F(x_b) \over {F(\infty)}}
	\int_{L_{{\rm min},b}} ^\infty dL L \bar \psi (B_0,L,z). 
\eeq
Note again the similarity of this expression to \eq\ (\ref{zmm'}).
As above, if we redefine the profile as the convolved ${\hat f}$
and include all resolved and unresolved  galaxies in the analysis,
we obtain the relation:
\beq\label{Iskyall}
        {\cal I}_{\rm sky}(>b,z)={1 \over {\Z4}} {d\chi \over dz}
        \int_{B_z} ^\infty dB_0
        \int_{\hat L_{{\rm min} ,b}} ^\infty dL L 
	 {F(\hat x_b) \over {F(\infty)}}\bar \psi (B_0,L,z).
\eeq
The differential distribution is obtained as 
$i_{\rm sky}(b, z)=-\partial {{\cal I}_{\rm sky}(>b,z)}/\partial b$, and 
the integrals of these over $z$ give the cumulative and differential 
total sky brightness as a function of $b$. Note that 
$i_{\rm sky}(b, z)=bf_{\rm sky}(b,z)$.

\subsection{Average Sky Brightness}

The ratio of the quantities described in above two tests gives the 
average \sb\ at or down to some \sb\ $b$.
This quantity as expected is independent of the cosmological model
parameters and depends only on the \sb\ profile and redshift and 
consequently, as already pointed out in P76, 
can be used to to determine the evolution of the 
surface brightness. In general, this
relation is more complicated than envisioned in P76 where
the discussion was aimed at the brightest cluster galaxies.
For a larger and varied sample of sources 
this relation is more complex and not as obvious. 
If, for purpose of illustration, we 
assume that $B_0$ and $r_0$ or $L$ are not correlated,
then from \eq\ (\ref{Fskyall}) and its counterpart for ${\cal I}$,
this ratio becomes
\beq\label{avskyb}
	<b(z, b)>=b \left(\int_{B_z} ^\infty F(x_b)h(B_0)dB_0
	/\int_{B_z} ^\infty x_b^2 f(x_b)h(B_0)dB_0\right),
\eeq
where $h(B_0)$ describes the distribution of the central surface brightness.
Note that for a delta function, or a relatively narrow distribution,
the above expression simplifies to $<b>/b=\eta(x_b)$ as is the 
case with individual galaxies. This demonstrates that the 
surface brightness profile, or the function $\eta$ based on it, plays 
a central role in cosmological studies of extended sources.

With minor modification
the above expressions will be valid for elliptical sources 
with constant ellipticity. For such sources 
with major and minor core radii $r_1$ and $r_2$,
the area within  any isophotal limit is proportional to
$\pi r_1 r_2$, so that if we define
$r_0=(r_1 r_2)^{1/2}$ the above expressions would apply, but we 
now have the additional integration over the possible dispersion
of the ellipticities or the ratio $r_1/r_2$.

\section{SUMMARY}

In this paper we deal with the analysis of the distribution of
redshift-size-flux (or magnitude) data on extrgalactic sources, 
in particular galaxies, which is often used for testing the evolution
of sources and/or the universe. The usual practice is to describe the 
characteristics and evolution of the galaxies in terms of a simple luminosity
function $\phi(L,z)$, as if the galaxies are point sources and the 
data is simply flux or magnitude limited.  We emphasize that in reality
these tests are more complicated.  The proper analysis
must involve the variation of the \sb\ profile and the multidimensional
distribution function of the parameters that describe this profile,
such as core radius, central surface brightness, luminosity, etc.;
$\psi(B_0, r_0, L, \alpha, ..., z)$. Neglecting these copmlexities
and the truncatios of the data produced by the \sb\ and angular size
limits can lead to grossly misleading results.

There are different ways one can account for these effects. The most 
efficient use of the observations comes from comparison of 
the full set of the data
with model predictions which include the effects of all biases that are 
encountered in the observational selection processes. 
Use of the isophotal values down to lowest posssible isophot and size
is a good example of this approach. We have described the correct analysis 
of such a data in terms of the multivariate luminosity function $\psi$.
We derive the relevant expressions, in terms of the \sb\
profile of spherical sources, which is to be compared with observations 
of the isophotal values of the fluxes, sizes and 
surface brightnesses. In general, because more than just the usual
flux (or magnitude) limit enters in the analysis these expressions are
relatively complex. Truncation of
the data due to other selection effects such as angular extent and \sb\
thresholds come in and  can play the dominant role
in defining the content of a sample of sources. The \sb\ profile of the 
galaxies plays a pivotal role in these calculations.

A second approach would be to select a subset of the data which yields to
a more straight forward comparison with models. For example, if we select
the more limited sample of large and well resolved sources, we can use
fluxes and sizes related to a metric (instead of isophotal) radius, such
as the radius defined in P76 for a constant value of the $\eta$ function.
We derive the relevant expressions for this case and show that they are
much simpler
than the isophotal ones, and resemble more closely  
the simple expressions for point sources. 
This method, therefore, would be more appropriate
for evaluation of the moments of the observed distributions in tests like the
redshift-flux or angular size-redshift tests. The isophotal method is
more appropriate approach for tests based on source counts and, 
in principle, can be used for samples of sources which include
resolved as well as unresolved sources. With the knowledge of the
PSF at hand one can use a modified \sb\ profile as convolution of 
the actual profile and the PSF. 

The above tests, aside from the complications in the modeling,
suffer from the additional shortcoming of needing elaborate
procedures for the analysis of the data. To overcome 
some of the difficulties in both of these areas
we propose a new method for the analysis and modeling of extended 
extragalactic sourses for the purpose of determining either
their evolution or the cosmological parameters.
This method is very well suited to modern digitized data,
and amounts to counting the numbers of pixels or summing their
intensity values. It is capable of using all the data as in the isophotal 
case but the expressions relating the data 
to models are  considerably simpler 
and are similar to the metric ones.
Their simplicity stems from the fact that they deal with
the surface brightness limit alone and do not include selection
based on  fluxes or sizes.
As already shown in P76, tests based on \sb\ tend to be 
more robust and simpler.

To reiterate, these new methods clearly have the following advantages.
	
a) The data analysis is obviously considerably simpler.

b) The expressions relating the observations to the distributions of the basic 
properties of galaxies is evidently simpler: Compare, e.g. 
\eqs\ (\ref{Next}) and (\ref{Fskyall}).

c) The dependence on the cosmological parameters is also 
considerably more straight forward: Instead of the dependence on the
volume $V(z)$, and on the luminosity and angular diameter distances,
$d_L$ and $d_A$, the new tests
depend primarily on the redshift and the much simpler line element $d\chi/dz$.
For example, for models with
zero cosmological constant this is equal to
$(c/H_0)(1+z)(\Omega z+1)^{-1/2}$,
where $\Omega$ is the density parameter and $H_0$ is the Hubble constant.

d) Because we are dealing with surface brightness, to first order the resulting
expressions are independent of the weak gravitational lensing effects due to
the inhomogeneities of the 
intervening matter distribution (clumpiness due to galaxies and clusters).
 
As mentioned above the modern digitized CCD data are ideally suited for
the task proposed here. However, 
several conditions are required for the proper
application of all the tests proposed here.
A good knowledge of the background sky
brightness due to all other sources except extragalactic sources 
underconsideration is needed
because we wish  to go to as low a surface brightness as possible.
Application of the new methods to integrated (over the redshift) values of 
${\cal F}_{\rm sky}$ and ${\cal I}_{\rm sky}$ to the whole data 
say, in a CCD frame, will require accurate flat fielding. However, for 
redshift dependent type analysis the requirements are similar to galaxy
based analysis, because redshifts are known for galaxies as a whole and 
not for indvidual pixels. For this kind of studies some identification of 
the pixels with galaxies is necessary so that we can assign redshifts
to pixels.
Secondly, when combining resolved and unresolved galaxies it is important that
contaminations due to other unresolved sources such as stars is kept 
to an acceptable value. And Finally, in all these tests one must deal
with several profiles; disks, spheroids and possibly a continuum of
superposition of both.
In future works we hope to use these new methods 
on the data from the Hubble Deep Field (see Williams et al., 1996)
as well as 
those from ground based observations.

This work was carried out while I  spent  a three month sabbatical leave at
the Space Telescope Science Institute. I would like to thank 
the director Robert Williams and the Staff of the Institute for their
support and hospitality. 
I also would like to thank Drs. H. Ferguson, M. Vogeley and M. Fall 
for valuable discussions on the general topic of this paper.

\end{document}